\documentclass[aip,jcp,reprint,groupedaddress]{revtex4-1}%
\usepackage{graphicx}
\usepackage{tabularx}
\usepackage{dcolumn}
\usepackage{longtable}
\usepackage{tensor}
\usepackage{color}
\usepackage{placeins}
\usepackage{bm}
\usepackage{amsmath}
\usepackage{amsfonts}
\usepackage{amssymb}%
\providecommand{\U}[1]{\protect\rule{.1in}{.1in}}
\providecommand{\U}[1]{\protect\rule{.1in}{.1in}}
\providecommand{\U}[1]{\protect\rule{.1in}{.1in}}
\begin{document}
\title{A time-reversible integrator for the time-dependent Schr\"{o}dinger equation
on an adaptive grid}
\author{Seonghoon Choi}
\author{Ji\v{r}\'{\i} Van\'{\i}\v{c}ek}
\email{jiri.vanicek@epfl.ch}
\affiliation{Laboratory of Theoretical Physical Chemistry, Institut des Sciences et
Ing\'enierie Chimiques, Ecole Polytechnique F\'ed\'erale de Lausanne (EPFL),
CH-1015, Lausanne, Switzerland}
\date{\today}

\begin{abstract}
One of the most accurate methods for solving the time-dependent
Schr\"{o}dinger equation uses a combination of the dynamic Fourier method with
the split-operator algorithm on a tensor-product grid. To reduce the number of
required grid points, we let the grid move together with the wavepacket, but
find that the na\"{\i}ve algorithm based on an alternate evolution of the
wavefunction and grid destroys the time reversibility of the exact evolution.
Yet, we show that the time reversibility is recovered if the wavefunction and
grid are evolved simultaneously during each kinetic or potential step; this is
achieved by using the Ehrenfest theorem together with the splitting method.
The proposed algorithm is conditionally stable, symmetric, time-reversible,
and conserves the norm of the wavefunction. The preservation of these
geometric properties is shown analytically and demonstrated numerically on a
three-dimensional harmonic model and collinear model of He--H$_{2}$
scattering. We also show that the proposed algorithm can be symmetrically
composed to obtain time-reversible integrators of an arbitrary even order. We
observed $10000$-fold speedup by using the tenth- instead of the second-order
method to obtain a solution with a time discretization error below $10^{-9}$.
Moreover, using the adaptive grid instead of the fixed grid resulted in a
64-fold reduction in the required number of grid points in the harmonic system
and made it possible to simulate the He--H$_{2}$ scattering for six times
longer, while maintaining reasonable accuracy.
Applicability of the algorithm to high-dimensional quantum
dynamics is demonstrated using the strongly anharmonic eight-dimensional H\'{e}non--Heiles model.

\end{abstract}
\maketitle

\graphicspath{{./figures/}{C:/Users/Jiri/Dropbox/Papers/Chemistry_papers/2019/Moving_grids/figures/}{"d:/Group Vanicek/Desktop/Choi/Moving_grid/figures/"}}

\section{\label{sec:intro}Introduction}

Understanding many dynamical phenomena in chemical physics requires the
solution of the time-dependent Schr\"{o}dinger
equation.\cite{book_Heller:2018, Engel_Zewail:1988, Kosloff:1988,
Stock_Woywod:1995, Begusic_Vanicek:2018, Ben-Nun_Martinez:2000,
Bircher_Rothlisberger:2017} This equation can be often solved both accurately
and efficiently by employing a combination of the dynamic Fourier method with
a high-order split-operator algorithm\cite{Feit_Steiger:1982,
book_Tannor:2007, book_Lubich:2008, book_Hairer_Wanner:2006,
Roulet_Vanicek:2019} on a tensor-product grid.\cite{Feit_Steiger:1982,
book_Tannor:2007, Kosloff_Kosloff:1983} However, for simulations that access
only a small portion of the tensor-product Hilbert space, more suitable
methods exist. These methods focus the available computational resources on
the important portions of the full Hilbert space.

The multiconfigurational time-dependent Hartree (MCTDH)
method\cite{Meyer_Cederbaum:1990, Manthe_Cederbaum:1992, Worth_Burghardt:2008}
and its multilayer extension\cite{Wang_Thoss:2003} reduce the required number
of basis functions by employing an optimized time-dependent basis set.
The recently developed pruned, collocation-based MCTDH approach\cite{Wodraszka_Carrington:2019} mitigates the exponential scaling of the MCTDH method with the number of dimensions by employing a pruned basis\cite{Wodraszka_Carrington:2016} and simultaneously extends the applicability of the MCTDH method to general potential energy surfaces by using a collocation grid.\cite{Wodraszka_Carrington:2018} Another way of reducing the required computational resources is by appropriately truncating
a lattice of Gaussian basis functions\cite{Davis_Heller:1979, Shimshovitz_Tannor:2012,
Halverson_Poirier:2012, Arai_Kono:2018} or a set of grid points.\cite{Pettey_Wyatt:2006, Lee_Chou:2018}
In particular, sparse-grid methods\cite{Avila_Carrington:2009, Gradinaru:2008, book_Lubich:2008,
Lauvergnat_Nauts:2014} reduce the number of required grid points, e.g., by employing the Smolyak
quadrature.\cite{Smolyak:1963} Making the grid
adaptive\cite{book_Thomson_Mastin:1997} is another common approach to reduce
the required number of grid points. For example, the adaptive moving grid has
been used to improve the quantum trajectory method\cite{Lopreore_Wyatt:1999,
Wyatt:1999} near wavefunction nodes,\cite{Wyatt_Bittner:2000, Wyatt:2002,
Hughes_Wyatt:2002} to treat the interaction of molecules with intense
time-dependent electromagnetic fields,\cite{Lu_Bandrauk:2001} and to reduce
the size\cite{Sim_Makri:1995, Adhikari_Billing:2000} of grids employed in
discrete variable representation (DVR),\cite{Lill_Light:1982, Light_Carrington:2007}
which has been widely employed to compute vibrational spectra.\cite{Bramley_Carrington:1993, Wang_Carrington:2009} The parallelized\cite{Khan_Adhikari:2013, Khan_Adhikari:2014} time-dependent
DVR\cite{Adhikari_Billing:2000, Puzari_Adhikari:2005} has enabled
quantum dynamics simulations in various multi-state and multi-dimensional systems.\cite{Mandal_Adhikari:2018}

To reduce the number of grid points and memory required for quantum
simulations of systems that occupy only a small part of the accessible
phase space at any given time, in this paper, we use an adaptive
tensor-product grid that moves according to the wavepacket expectation values of position and
momentum. The most na\"{\i}ve approach is to evolve the
grid only after each time step of the wavefunction propagation. However, this
na\"{\i}ve grid adaptation breaks the symmetry and, therefore, also the time
reversibility of the time propagation scheme.

We find that the time reversibility is recovered when the grid is evolved
simultaneously with the wavepacket. The resulting algorithm is not only
symmetric and time-reversible, but also norm-preserving and conditionally
stable (i.e., stable for small enough time steps). In addition, because of its
symmetry, this algorithm can be composed by various symmetric composition
schemes to obtain higher-order integrators.\cite{book_Hairer_Wanner:2006,
Suzuki:1990, Yoshida:1990, Kahan_Li:1997, Sofroniou_Spaletta:2005,
Roulet_Vanicek:2019}
As well as having favorable geometric properties, the proposed algorithm is also very simple to
implement because its implementation does not depend on the form of the potential energy function and because
the grid adaptation requires no adjustable parameters.

The remainder of this paper is organized as follows: In Sec.~\ref{sec:theory},
we give a brief overview of the split-operator algorithm and dynamic Fourier
method, including their discretized implementation on a tensor-product grid.
Then, we demonstrate the breakdown of the time reversibility by the na\"{\i}ve
grid adaptation, and the recovery of the time reversibility by employing a
combination of the Ehrenfest theorem\cite{Ehrenfest:1927} and splitting
method.\cite{book_Hairer_Wanner:2006, book_Lubich:2008} In
Sec.~\ref{sec:numerical_examples}, we numerically confirm the geometric and
convergence properties of the proposed algorithm using a three-dimensional
harmonic model of electronic excitation and a two-dimensional modified
Secrest--Johnson\cite{Secrest_Johnson:1966, Campos-Martinez_Coalson:1990}
model of He--H$_{2}$ scattering.
Using the highly nonlinear eight-dimensional H\'{e}non--Heiles system, we demonstrate that the
proposed algorithm can be also used for high-dimensional quantum dynamics, which is beyond the reach
of the conventional split-operator algorithm on a fixed tensor-product grid.
Section~\ref{sec:conclusion} concludes the paper.

\section{\label{sec:theory}Theory}

The time-dependent Schr\"{o}dinger equation,
\begin{equation}
\frac{d |\psi_{t}\rangle}{dt} = - \frac{i}{\hbar} \hat{H} |\psi_{t}\rangle,
\label{eq:tdse}%
\end{equation}
where $\hat{H}$ is a time-independent Hamiltonian and $| \psi_{t} \rangle$ is
the quantum state at time $t$, has the solution $| \psi_{t} \rangle= \hat
{U}(t) |\psi_{0} \rangle$ with the exact evolution operator $\hat{U}(t) :=
e^{-it\hat{H}/\hbar}$. In general, this exact solution must be approximated by
one of many possible time propagation schemes. Here, we will only discuss the
split-operator algorithms\cite{Feit_Steiger:1982, book_Tannor:2007,
book_Lubich:2008} in detail because the splitting
method\cite{book_Hairer_Wanner:2006, book_Lubich:2008} is crucial for the
time-reversible grid adaptation that we derive in
Sec.~\ref{subsec:recovery_of_symmetry_and_time_reversibility}.

\subsection{\label{subsec:split-operator} Split-operator algorithm and dynamic
Fourier method}

The splitting method requires the Hamiltonian to be separable into a sum of
kinetic and potential energy operators:
\begin{equation}
\hat{H} = T(\hat{\vec{p}}\,) + V(\hat{\vec{q}}\,), \label{eq:hamiltonian}%
\end{equation}
where $\vec{p}$ and $\vec{q}$ are $D$-dimensional momentum and position,
respectively. For orthogonal coordinates, the kinetic energy $T(\vec{p}\,)$
has a simple form
\begin{equation}
T(\vec{p}) = \frac{1}{2} \vec{p}\,^{T} \cdot m^{-1} \cdot\vec{p},
\label{eq:KE_operator}%
\end{equation}
where $m$ is a real symmetric (and often diagonal) $D \times D$ mass matrix.
For Hamiltonians of form~(\ref{eq:hamiltonian}), all split-operator algorithms
can be expressed as a composition\cite{book_Hairer_Wanner:2006} of kinetic
[$\hat{U}_{\hat{T}}(t)$] and potential [$\hat{U}_{\hat{V}}(t)$] evolution
operators, where $\hat{U}_{\hat{A}}(t):= e^{-i t \hat{A}/\hbar}$ is the exact
evolution operator for a Hamiltonian $\hat{H} = \hat{A}$. The simplest
first-order split-operator algorithm\cite{Trotter:1959} has the approximate
evolution operator
\begin{equation}
\hat{U}_{\text{TV}}(\Delta t) := \hat{U}_{\hat{T}}(\Delta t) \hat{U}_{\hat{V}%
}(\Delta t). \label{eq:SO_TV}%
\end{equation}
This ``TV'' algorithm can be composed with its adjoint, $\hat{U}_{\text{VT}%
}(\Delta t) := \hat{U}_{\text{TV}} (-\Delta t)^{-1}$, to obtain the
second-order TVT algorithm,\cite{Strang:1968}
\begin{align}
\hat{U}_{\text{TVT}}(\Delta t)  &  := \hat{U}_{\text{TV}}(\Delta t/2) \hat
{U}_{\text{VT}}(\Delta t/2)\nonumber\\
&  = \hat{U}_{\hat{T}}(\Delta t/2) \hat{U}_{\hat{V}}(\Delta t) \hat{U}%
_{\hat{T}} (\Delta t/2), \label{eq:SO_TVT}%
\end{align}
which is symmetric, i.e., satisfies $\hat{U}_{\text{TVT}}(\Delta t) = \hat
{U}_{\text{TVT}}(- \Delta t)^{-1}$. Because it is symmetric, the TVT algorithm
can be recursively composed by symmetric
schemes\cite{book_Lubich:2008,book_Hairer_Wanner:2006,Suzuki:1990,Yoshida:1990,
Kahan_Li:1997, Sofroniou_Spaletta:2005} to obtain symmetric algorithms
\begin{equation}
\hat{U}_{\text{comp}}(\Delta t) := \hat{U}_{\text{TVT}}(\gamma_{N_{\text{comp}%
}} \Delta t) \cdots\hat{U}_{\text{TVT}}(\gamma_{1} \Delta t)
\label{eq:comp_TVT}%
\end{equation}
of arbitrary even orders of accuracy in the time step, where $\gamma_{k}$ is
the $k$th composition coefficient (with $\sum_{k=1}^{N_{\text{comp}}}
\gamma_{k} =1$, $\gamma_{N_{\text{comp}}+1-k} = \gamma_{k}$) and
$N_{\text{comp}}$ is the number of composition steps. [Exchanging $V$ and $T$
in algorithms~(\ref{eq:SO_TV})--(\ref{eq:comp_TVT}) results in another set of
split-operator algorithms.] The higher-order algorithms, despite their higher
computational cost per time step, are often more efficient if high accuracy is
desired. Moreover, all symmetric split-operator algorithms are examples of
geometric integrators because they preserve many important geometric
properties of the exact solution of Eq.~(\ref{eq:tdse}), namely linearity,
unitarity, symplecticity, stability, symmetry, and time
reversibility.\cite{book_Hairer_Wanner:2006,book_Lubich:2008,book_Leimkuhler_Reich:2004}
For additional details about the properties and numerical implementation of
the higher-order split-operator algorithms, we refer the reader to Ref.~\onlinecite{Roulet_Vanicek:2019}.

The action of compositions of $\hat{U}_{\hat{T}}(\Delta t)$ and $\hat{U}%
_{\hat{V}}(\Delta t)$ on $|\psi\rangle$ is simply evaluated using the dynamic
Fourier method\cite{Feit_Steiger:1982, Kosloff_Kosloff:1983, book_Tannor:2007}
in which the action of a function $g(\hat{\vec{x}})$ of an operator $\hat
{\vec{x}}$ on the wavepacket $| \psi\rangle$ is evaluated as $g(\vec{x})
\psi(\vec{x})$ in the $\vec{x}$-representation, where $\vec{x}$ is either the
position $\vec{q}$ or momentum $\vec{p}$. The wavefunction is transformed, if
required, to the $\vec{x}$-representation by either the Fourier or inverse
Fourier transformation:
\begin{align}
\tilde{\psi}(\vec{p})  &  = (2 \pi\hbar)^{-D/2} \int\psi(\vec{q}) e^{- i
\vec{p} \cdot\vec{q} / \hbar} d^{D}q,\label{eq:wavepacket_q_to_p_nd}\\
\psi(\vec{q})  &  = (2 \pi\hbar)^{-D/2} \int\tilde{\psi}(\vec{p}) e^{i \vec{p}
\cdot\vec{q} / \hbar} d^{D}p, \label{eq:wavepacket_p_to_q_nd}%
\end{align}
where $\vec{p} \cdot\vec{q} := \sum_{l=1}^{D} p_{l} q_{l}$.

\subsection{\label{subsec:Fourier_method_grid} Dynamic Fourier method on a
grid}

The dynamic Fourier method on a grid follows the same approach as in
Sec.~\ref{subsec:split-operator} except that $g(\vec{x})\psi(\vec{x})$ and the
integral transforms~(\ref{eq:wavepacket_q_to_p_nd}) and
(\ref{eq:wavepacket_p_to_q_nd}) are now discretized on a grid, consisting of
points $\vec{x}^{I}$ for all $I \in\mathcal{I}$. Here, the multi-index $I =
(i_{1}, \dots, i_{D})$ is an ordered $D$-tuple of integers from the set
$\mathcal{I}$ of all admissible multi-indices, where
\begin{align}
\mathcal{I} := \{ (i_{1}, \dots, i_{D}): i_{l} \in\{ 0, \dots, N_{l} - 1
\}\nonumber\\
\text{for all} \quad l \in\{1, \dots, D \} \}, \label{eq:multi_index_I}%
\end{align}
and $N_{l}$ is the number of grid points in the $l$th dimension. When
iterating over all admissible multi-indices, we will simply write $I
\in\mathcal{I}$. The $D$ coordinates of the grid point $\vec{x}^{I}$ are given
by
\begin{equation}
x_{l}^{I} := x_{\text{ctr}, l} + (i_{l} - N_{l}/2) \Delta x_{l} \quad
\text{for} \quad l \in\{1, \dots, D \}, \label{eq:x_I_definition}%
\end{equation}
where $\vec{x}_{\text{ctr}}$ is the $\vec{x}$-grid center and $\Delta\vec{x}$
are the $\vec{x}$-spacings of the grid. Note that $\Delta q_{l} \Delta p_{l} =
2 \pi\hbar/N_{l}$.

Application of the operator $g(\hat{\vec{x}})$ to $\psi$ in $\vec{x}%
$-representation is given by
\begin{equation}
g(\vec{x})\psi(\vec{x})\overset{\text{grid}}{=}g(\vec{x}^{I})\psi(\vec{x}%
^{I}),\label{eq:g_x_on_psi}%
\end{equation}
where $\overset{\text{grid}}{=}$ denotes \textquotedblleft is represented on a
grid as.\textquotedblright\ The integral transforms
(\ref{eq:wavepacket_q_to_p_nd}) and (\ref{eq:wavepacket_p_to_q_nd}) are
discretized on a grid as
\begin{align}
\tilde{\psi}(\vec{p}\,^{K}) &  =C_{q}\sum_{J\in\mathcal{I}}\psi(\vec{q}%
\,^{J})e^{-i\vec{p}\,^{K}\cdot\,\vec{q}\,^{J}/\hbar},\quad K\in\mathcal{I}%
,\label{eq:fourier_pos_to_mom}\\
\psi(\vec{q}\,^{J}) &  =C_{p}\sum_{K\in\mathcal{I}}\tilde{\psi}(\vec{p}%
\,^{K})e^{i\vec{p}\,^{K}\cdot\,\vec{q}\,^{J}/\hbar},\quad J\in\mathcal{I}%
,\label{eq:fourier_mom_to_pos}%
\end{align}
with prefactors $C_{x}:=\prod_{l=1}^{D}(\Delta x_{l}/\sqrt{2\pi\hbar})$. To
express Eqs.~(\ref{eq:fourier_pos_to_mom}) and (\ref{eq:fourier_mom_to_pos})
in terms of the standard discrete Fourier transform (DFT), we scale the
wavefunctions as $\tilde{\psi}^{K}:=\tilde{\psi}(\vec{p}\,^{K})/\sqrt{C_{q}}$,
$\psi^{J}:=\psi(\vec{q}\,^{J})/\sqrt{C_{p}}$ and use
Eq.~(\ref{eq:x_I_definition}). As a result, we obtain
\begin{align}
\tilde{\psi}^{K} &  =\frac{1}{\sqrt{N}}\sum_{J\in\mathcal{I}}e^{-2\pi i\langle
K,J\rangle}e^{-it_{KJ}/\hbar}\psi^{J},\quad K\in\mathcal{I}%
,\label{eq:fourier_pos_to_mom_b}\\
\psi^{J} &  =\frac{1}{\sqrt{N}}\sum_{K\in\mathcal{I}}e^{2\pi i\langle
J,K\rangle}e^{it_{KJ}/\hbar}\tilde{\psi}^{K},\quad J\in\mathcal{I}%
,\label{eq:fourier_mom_to_pos_b}%
\end{align}
where $N:=\prod_{l=1}^{D}N_{l}$ denotes the total number of grid points,
\begin{align}
t_{KJ} &  =\sum_{l=1}^{D}[(p_{\text{ctr},l}-\Delta p_{l}N_{l}/2)(q_{\text{ctr}%
,l}-\Delta q_{l}N_{l}/2)\nonumber\\
&  \quad+p_{\text{ctr},l}j_{l}\Delta q_{l}+q_{\text{ctr},l}k_{l}\Delta
p_{l}-\pi\hbar(j_{l}+k_{l})],\label{eq:t_definition}%
\end{align}
and the multi-index inner product $\langle K,J\rangle:=\sum_{l=1}^{D}%
k_{l}j_{l}/N_{l}$. The scaled wavefunction $\tilde{\psi}^{K}$ can be viewed as
a standard DFT of $e^{-it_{KJ}/\hbar}\psi^{J}$ and $\psi^{J}$ as a standard
inverse DFT of $e^{it_{KJ}/\hbar}\tilde{\psi}^{K}$. In practice, the DFT is
implemented using the celebrated fast Fourier transform algorithm,
which has been parallelized via the message passing interface (MPI) and the open
multi-processing (openMP) interface.\cite{Frigo_Johnson:2005}

\subsection{\label{subsec:shifting_grid}Shifting the grid}

In Sec.~\ref{subsec:Fourier_method_grid}, we assumed the grid centers to be
fixed, but this assumption will now be dropped to allow for grid shifting.
Moreover, all quantities defined in Sec.~\ref{subsec:Fourier_method_grid} will
be re-expressed in a more compact matrix form.
Equations~(\ref{eq:fourier_pos_to_mom_b}) and (\ref{eq:fourier_mom_to_pos_b})
thus become
\begin{align}
\tilde{\boldsymbol{\psi}}(\vec{p}_{\text{ctr}})  &  = \mathbf{f}(\vec
{q}_{\text{ctr}}, \vec{p}_{\text{ctr}}) \boldsymbol{\psi}(\vec{q}_{\text{ctr}%
}),\label{eq:fourier_pos_to_mom_c}\\
\boldsymbol{\psi}(\vec{q}_{\text{ctr}})  &  = \tilde{\mathbf{f}}(\vec
{q}_{\text{ctr}}, \vec{p}_{\text{ctr}}) \tilde{\boldsymbol{\psi}}(\vec
{p}_{\text{ctr}}), \label{eq:fourier_mom_to_pos_c}%
\end{align}
where the ``vectors'' $\boldsymbol{\psi}(\vec{q}_{\text{ctr}})$ and
$\tilde{\boldsymbol{\psi}}(\vec{p}_{\text{ctr}})$ are rank-$D$ tensors with
$N$ components $\psi^{J}(\vec{q}_{\text{ctr}})$ and $\tilde{\psi}^{K}(\vec
{p}_{\text{ctr}})$, respectively, and the ``matrices'' representing the
Fourier transforms are
\begin{align}
&  [\mathbf{f}(\vec{q}_{\text{ctr}}, \vec{p}_{\text{ctr}})]_{KJ} := \frac
{1}{\sqrt{N}} e^{-2 \pi i \langle K, J \rangle} e^{-i t_{KJ}(\vec
{q}_{\text{ctr}}, \vec{p}_{\text{ctr}})/\hbar},\label{eq:F_defintion}\\
&  \tilde{\mathbf{f}}(\vec{q}_{\text{ctr}}, \vec{p}_{\text{ctr}}) =
\mathbf{f}(\vec{q}_{\text{ctr}}, \vec{p}_{\text{ctr}})^{-1} = \mathbf{f}%
(\vec{q}_{\text{ctr}}, \vec{p}_{\text{ctr}})^{\dagger}.
\label{eq:F_inv_definition}%
\end{align}
In Eqs.~(\ref{eq:fourier_pos_to_mom_c}) and (\ref{eq:fourier_mom_to_pos_c}), a
compact notation for the ``matrix-vector'' multiplication, defined by
$(\mathbf{a} \boldsymbol{\psi})_{K} := \sum_{J\in\mathcal{I}} a_{KJ} \psi^{J}%
$, $K \in\mathcal{I}$, is employed, and $\,\tilde{}\,$ above a matrix denotes
that it is applied to a wavefunction in the $\vec{p}$-representation.

Equation~(\ref{eq:fourier_mom_to_pos_c}) expresses that the momentum
wavefunction $\tilde{\boldsymbol{\psi}}(\vec{p}_{\text{ctr}})$ represented on
the $\vec{p}$-grid centered at $\vec{p}_{\text{ctr}}$ is transformed to the
position wavefunction $\boldsymbol{\psi}(\vec{q}_{\text{ctr}})$ represented on
the $\vec{q}$-grid centered at $\vec{q}_{\text{ctr}}$ by applying
$\tilde{\mathbf{f}}(\vec{q}_{\text{ctr}}, \vec{p}_{\text{ctr}})$. Similarly,
$\boldsymbol{\psi}(\vec{q}_{\text{ctr}})$ is transformed to $\tilde
{\boldsymbol{\psi}}(\vec{p}_{\text{ctr}})$ by applying $\mathbf{f}(\vec
{q}_{\text{ctr}}, \vec{p}_{\text{ctr}})$ according to
Eq.~(\ref{eq:fourier_pos_to_mom_c}).

\subsection{\label{subsec:so_algorithm_grid}Split-operator algorithm on a
grid}

Kinetic and potential evolution operators $\hat{U}_{\hat{T}}(\Delta t)$ and
$\hat{U}_{\hat{V}} (\Delta t)$, which are composed to obtain any
split-operator algorithm (see Sec.~\ref{subsec:split-operator}), are
discretized on a grid as diagonal finite-dimensional tensors
\begin{align}
[\mathbf{U}_{\text{V}}(\Delta t, \vec{q}_{\text{ctr}})]_{JJ^{\prime}}  &  =
\delta_{JJ^{\prime}} e^{- i \Delta t V\boldsymbol{(}\vec{q}\,^{J}(\vec
{q}_{\text{ctr}})\boldsymbol{)}/\hbar},\label{eq:U_V_tensor}\\
[\tilde{\mathbf{U}}_{\text{T}}(\Delta t, \vec{p}_{\text{ctr}})]_{KK^{\prime}}
&  = \delta_{KK^{\prime}} e^{- i \Delta t T\boldsymbol{(}\vec{p}\,^{K}(\vec
{p}_{\text{ctr}})\boldsymbol{)}/\hbar}. \label{eq:U_T_tensor}%
\end{align}
Therefore, the time-evolved wavefunctions are
\begin{align}
\langle\vec{q} \, | \hat{U}_{\hat{V}}(\Delta t) | \psi\rangle &
\overset{\text{grid}}{=} \sqrt{C_{p}} \mathbf{U}_{\text{V}}(\Delta t, \vec
{q}_{\text{ctr}}) \boldsymbol{\psi}(\vec{q}_{\text{ctr}}),\\
\langle\vec{p}\, | \hat{U}_{\hat{T}}(\Delta t) | \psi\rangle &
\overset{\text{grid}}{=} \sqrt{C_{q}} \tilde{\mathbf{U}}_{\text{T}}(\Delta t,
\vec{p}_{\text{ctr}}) \tilde{\boldsymbol{\psi}} (\vec{p}_{\text{ctr}});
\end{align}
note that the scaled wavefunctions, $\psi^{J}$ and $\tilde{\psi}^{K}$, must be
scaled back to $\psi(\vec{q}\,^{J})$ and $\tilde{\psi}(\vec{p}\,^{K})$ at the
end of the propagation with factors $\sqrt{C_{p}}$ and $\sqrt{C_{q}}$,
respectively (see Sec~\ref{subsec:Fourier_method_grid}).

\subsection{\label{subsec:loss_of_geom_prop} Loss of linearity by the grid
adaptation}

To be specific, we now assume that the initial wavefunction is provided in the
$\vec{q}$-representation and that the solution at time $t$ is desired also in
the $\vec{q}$-representation. Let the adaptive $\vec{x}$-grid be centered at
the wavefunction's $\vec{x}$-expectation value. The resulting equations of
motion for the wavefunction and grid centers, $\vec{x}_{t} := \vec
{x}_{\text{ctr}}(t)$, are
\begin{align}
\dot{\boldsymbol{\psi}}_{t}(\vec{q}_{t})  &  = - \frac{i}{\hbar}
\mathbf{H}(\vec{q}_{t}, \vec{p}_{t}) \boldsymbol{\psi}_{t}(\vec{q}%
_{t}),\label{eq:eom_wf_case_I}\\
\vec{q}_{t}  &  = \langle\vec{\mathbf{q}}(\vec{q}_{t}) \rangle
_{\boldsymbol{\psi}_{t} (\vec{q}_{t})},\label{eq:eom_qt_case_I}\\
\vec{p}_{t}  &  = \langle\tilde{\vec{\mathbf{p}}}(\vec{p}_{t}) \rangle
_{\tilde{\boldsymbol{\psi}}_{t} (\vec{p}_{t}),} \label{eq:eom_pt_case_I}%
\end{align}
where $\tilde{\boldsymbol{\psi}}_{t}(\vec{p}_{t}) = \mathbf{f}(\vec{q}_{t},
\vec{p}_{t}) \boldsymbol{\psi}_{t}(\vec{q}_{t})$, $\mathbf{H}(\vec{q}_{t},
\vec{p}_{t})$ is the Hamiltonian in the $\vec{q}$-representation, containing
appropriate Fourier transforms, and represented on the grid centered at
$\vec{x}_{t}$, $[\vec{\mathbf{x}}(\vec{x}_{t})]_{II^{\prime}} :=
\delta_{II^{\prime}} \vec{x}^{I}(\vec{x}_{t})$, and $\langle\mathbf{O}(\vec
{x}_{t}) \rangle_{\boldsymbol{\psi}_{t}(\vec{x}_{t})} := \langle
\boldsymbol{\psi}_{t}(\vec{x}_{t}) | \mathbf{O}(\vec{x}_{t}) \boldsymbol{\psi
}_{t}(\vec{x}_{t}) \rangle$; the inner product between $\boldsymbol{\psi}$ and
$\boldsymbol{\phi}$ is defined as
\begin{equation}
\langle\boldsymbol{\psi}(\vec{x}_{t}) | \boldsymbol{\phi}(\vec{x}_{t})
\rangle:= \left(  \prod_{l=1}^{D} \Delta x_{l} \right)  \sum_{I \in
\mathcal{I}} \psi\boldsymbol{(}\vec{x}^{I}(\vec{x}_{t})\boldsymbol{)}^{\ast}
\phi\boldsymbol{(}\vec{x}^{I}(\vec{x}_{t})\boldsymbol{)}.
\label{eq:scalar_prod}%
\end{equation}

The grid adaptation leads to the loss of some geometric properties even if
Eqs.~(\ref{eq:eom_wf_case_I})--(\ref{eq:eom_pt_case_I}) are solved exactly. In
Eq.~(\ref{eq:eom_wf_case_I}), the Hamiltonian is nonlinear due to its
dependence on $\psi_{t}$ (via $\vec{q}_{t}$ and $\vec{p}_{t}$); the
corresponding evolution operator is, therefore, also nonlinear, and does not
preserve the inner product.\cite{book_Halmos:1942} As a consequence, the
symplectic two-form\cite{book_Lubich:2008} $\omega(\boldsymbol{\psi},
\boldsymbol{\phi}) := -2 \hbar\mathrm{Im} \langle\boldsymbol{\psi} |
\boldsymbol{\phi} \rangle$ is not preserved, either. In contrast, the exact
solution of Eqs.~(\ref{eq:eom_wf_case_I})--(\ref{eq:eom_pt_case_I}) does
preserve the norm and is both symmetric and time-reversible.

\subsection{\label{subsec:loss_of_geom_prop_naive} Loss of time reversibility
by the na\"{\i}ve adaptive grid}

Due to their mutual coupling, Eqs.~(\ref{eq:eom_wf_case_I}%
)--(\ref{eq:eom_pt_case_I}) cannot be, in general, solved analytically. The
na\"{\i}ve adaptive grid approximation decouples the equations for the
wavefunction and grid evolutions by first solving Eq.~(\ref{eq:eom_wf_case_I})
for $\boldsymbol{\psi}_{t}$ with fixed $\vec{q}_{t}$ and $\vec{p}_{t}$ during
the time $0 \leq t \leq\Delta t$, obtaining $\boldsymbol{\psi}_{\Delta t}%
(\vec{q}_{0}) = \mathbf{U}(\Delta t;\vec{q}_{0}, \vec{p}_{0}) \boldsymbol{\psi
}_{0}(\vec{q}_{0})$, where
\begin{equation}
\mathbf{U}(\Delta t; \vec{q}_{0}, \vec{p}_{0}) := e^{- i \Delta t
\mathbf{H}(\vec{q}_{0}, \vec{p}_{0})/\hbar}. \label{eq:naive_wf_solution}%
\end{equation}
[In practice, $\mathbf{U}(\Delta t;\vec{q}_{0}, \vec{p}_{0}) \boldsymbol{\psi}_{0}(\vec{q}_{0})$ is approximated numerically using time propagation schemes, such as the split-operator algorithm, short iterative Lanczos scheme,\cite{Lanczos:1950,Park_Light:1986} or Crank--Nicolson\cite{Crank_Nicolson:1947,McCullough_Wyatt:1971} method.]
The grid centers are then updated using the propagated wavefunction
$\boldsymbol{\psi}_{\Delta t}(\vec{q}_{0})$:
\begin{align}
\vec{q}_{\Delta t}  &  = \langle\vec{\mathbf{q}}(\vec{q}_{0}) \rangle
_{\boldsymbol{\psi}_{\Delta t}(\vec{q}_{0})},\\
\vec{p}_{\Delta t}  &  = \langle\tilde{\vec{\mathbf{p}}}(\vec{p}_{0})
\rangle_{\tilde{\boldsymbol{\psi}}_{\Delta t}(\vec{p}_{0})}.
\end{align}
Finally, the wavefunction $\boldsymbol{\psi}_{\Delta t}(\vec{q}_{0})$ is
represented on the updated grid:
\begin{align}
\tilde{\boldsymbol{\psi}}_{\Delta t}(\vec{p}_{\Delta t})  &  = \mathbf{f}%
(\vec{q}_{0}, \vec{p}_{\Delta t}) \boldsymbol{\psi}_{\Delta t}(\vec{q}_{0}),\\
\boldsymbol{\psi}_{ \Delta t}(\vec{q}_{ \Delta t})  &  = \tilde{\mathbf{f}%
}(\vec{q}_{ \Delta t}, \vec{p}_{ \Delta t}) \tilde{\boldsymbol{\psi}}_{ \Delta
t}(\vec{p}_{ \Delta t}).
\end{align}
The overall evolution operator for the na\"{\i}ve adaptive grid is,
therefore,
\begin{align}
\mathbf{U}  &  _{\text{na\"{\i}ve}}(\Delta t; \boldsymbol{\psi}_{0}, \vec
{q}_{0}, \vec{p}_{0})\nonumber\\
&  := \tilde{\mathbf{f}}(\vec{q}_{\Delta t}, \vec{p}_{\Delta t})\mathbf{f}%
(\vec{q}_{0}, \vec{p}_{\Delta t}) \mathbf{U}(\Delta t;\vec{q}_{0}, \vec{p}%
_{0}),
\end{align}
where the dependence of $\mathbf{U}_{\text{na\"{\i}ve}}$ on $\boldsymbol{\psi
}_{0}$ comes from the dependence of $\vec{q}_{\Delta t}$ and $\vec{p}_{\Delta
t}$ on $\boldsymbol{\psi}_{0}$.

The time propagation on the na\"{\i}ve adaptive grid preserves the norm $\|
\boldsymbol{\psi}_{t}(\vec{q}_{t}) \| := \langle\boldsymbol{\psi}_{t}(\vec
{q}_{t})|\boldsymbol{\psi}_{t}(\vec{q}_{t}) \rangle^{1/2}$ because
$\mathbf{U}_{\text{na\"{\i}ve}}$ is a composition of three norm-preserving
operators: That $\mathbf{f}$ and $\tilde{\mathbf{f}}$ preserve the norm
follows from Eq.~(\ref{eq:F_inv_definition}), and $\mathbf{U}(\Delta t,
\vec{q}_{0}, \vec{p}_{0})$ preserves the norm because $\mathbf{U}^{\dagger} =
\mathbf{U}^{-1}$ [see Eq.~(\ref{eq:naive_wf_solution})].

A symmetric operator is time-reversible [i.e., satisfies $\mathbf{U}(-\Delta
t) \mathbf{U}(\Delta t) = 1$] and vice versa. Both of these properties are
lost in the na\"{\i}ve adaptive grid approach because
\begin{align}
&  \mathbf{U}_{\text{na\"{\i}ve}}(-\Delta t; \boldsymbol{\psi}_{\Delta t},
\vec{q}_{\Delta t}, \vec{p}_{\Delta t}) \mathbf{U}_{\text{na\"{\i}ve}}(\Delta
t; \boldsymbol{\psi}_{0}, \vec{q}_{0}, \vec{p}_{0})\nonumber\\
&  = \tilde{\mathbf{f}}(\vec{q}_{0}\,^{\prime}, \vec{p}_{0}\,^{\prime})
\mathbf{f}(\vec{q}_{\Delta t}, \vec{p}_{0}\,^{\prime})\mathbf{U}(-\Delta t;
\vec{q}_{\Delta t}, \vec{p}_{\Delta t})\nonumber\\
&  \times\tilde{\mathbf{f}}(\vec{q}_{\Delta t}, \vec{p}_{\Delta t}%
)\mathbf{f}(\vec{q}_{0}, \vec{p}_{\Delta t})\mathbf{U}(\Delta t; \vec{q}_{0},
\vec{p}_{0})\neq1, \label{eq:naive_t_rev_fail}%
\end{align}
where
\begin{align}
\vec{q}_{0}\,^{\prime}  &  = \langle\vec{\mathbf{q}}(\vec{q}_{\Delta t})
\rangle_{\boldsymbol{\psi}_{0}(\vec{q}_{\Delta t})} \neq\vec{q}_{0},\\
\vec{p}_{0}\,^{\prime}  &  = \langle\tilde{\vec{\mathbf{p}}}(\vec{p}_{\Delta
t}) \rangle_{\tilde{\boldsymbol{\psi}}_{0}(\vec{p}_{\Delta t})} \neq\vec
{p}_{0},\\
\boldsymbol{\psi}_{0} (\vec{q}_{\Delta t})  &  = \mathbf{U}(-\Delta t; \vec
{q}_{\Delta t}, \vec{p}_{\Delta t}) \boldsymbol{\psi}_{\Delta t}(\vec
{q}_{\Delta t}),\\
\tilde{\boldsymbol{\psi}}_{0}(\vec{p}_{\Delta t})  &  = \mathbf{f}(\vec
{q}_{\Delta t}, \vec{p}_{\Delta t}) \boldsymbol{\psi}_{0}(\vec{q}_{\Delta t}).
\end{align}
Note that inequality~(\ref{eq:naive_t_rev_fail}) would still hold even in the
unlikely situation that, by chance, $\vec{q}_{0}\,^{\prime} = \vec{q}_{0}$ and
$\vec{p}_{0}\,^{\prime} = \vec{p}_{0}$. As we shall see below, to preserve the
symmetry and time reversibility, the grid must be evolved simultaneously with
the wavefunction.

\subsection{\label{subsec:recovery_of_symmetry_and_time_reversibility}Recovery
of time reversibility by a combination of the splitting method and Ehrenfest
theorem}

The Ehrenfest theorem\cite{Ehrenfest:1927} states that the time derivatives of
the position and momentum expectation values satisfy
\begin{align}
\dot{\vec{q}}_{t}  &  = \left\langle \partial\mathbf{H}/\partial\vec{p}
(\vec{q}_{t}, \vec{p}_{t}) \right\rangle _{\boldsymbol{\psi}_{t}(\vec{q}_{t}%
)},\label{eq:eom_qt_case_II}\\
\dot{\vec{p}}_{t}  &  = - \left\langle \partial\mathbf{H}/\partial\vec{q}%
(\vec{q}_{t}, \vec{p}_{t}) \right\rangle _{\boldsymbol{\psi}_{t} (\vec{q}%
_{t})}. \label{eq:eom_pt_case_II}%
\end{align}
The system of differential and algebraic Eqs.~(\ref{eq:eom_wf_case_I}),
(\ref{eq:eom_qt_case_I}), (\ref{eq:eom_pt_case_I}) for $\boldsymbol{\psi}_{t},
\vec{q}_{t}, \vec{p}_{t}$ is equivalent to and, hence, can be replaced with
the system of differential Eqs.~(\ref{eq:eom_wf_case_I}),
(\ref{eq:eom_qt_case_II}), (\ref{eq:eom_pt_case_II}). These equations can be
solved analytically if $\hat{H} = V(\hat{\vec{q}})$ or $\hat{H} = T(\hat
{\vec{p}})$. This is the essence of the splitting
method\cite{book_Hairer_Wanner:2006, book_Lubich:2008} (see
Sec.~\ref{subsec:split-operator}).

\subsubsection{\label{subsubsec:pot_prop}Potential propagation: $\hat{H} =
V(\hat{\vec{q}}\,)$}

When $\hat{H} = V(\hat{\vec{q}}\,)$, Eqs.~(\ref{eq:eom_wf_case_I}),
(\ref{eq:eom_qt_case_II}), (\ref{eq:eom_pt_case_II}) become
\begin{align}
\dot{\boldsymbol{\psi}}_{t}(\vec{q}_{t})  &  = -\frac{i}{\hbar} \mathbf{V}%
(\vec{q}_{t}) \boldsymbol{\psi}_{t}(\vec{q}_{t}), \label{eq:eom_wf_case_III_V}%
\\
\dot{\vec{q}}_{t}  &  = 0,\label{eq:eom_qt_case_III_V}\\
\dot{\vec{p}}_{t}  &  = - \left\langle \partial\mathbf{V}/\partial\vec{q}
(\vec{q}_{t}) \right\rangle _{\boldsymbol{\psi}_{t}(\vec{q}_{t})},
\label{eq:eom_pt_case_III_V}%
\end{align}
where $[\mathbf{V}(\vec{q}_{t})]_{JJ^{\prime}} = \delta_{JJ^{\prime}}
V\boldsymbol{(}\vec{q}\,^{J}(\vec{q}_{t})\boldsymbol{)}$. These equations have
an exact analytical solution for arbitrarily long time $t$, namely:
\begin{align}
\boldsymbol{\psi}_{t}(\vec{q}_{t})  &  = \mathbf{U}_{\text{V}}(t, \vec{q}_{0})
\boldsymbol{\psi}_{0}(\vec{q}_{0}),\label{eq:eom_wf_solution_case_III_V}\\
\vec{q}_{t}  &  = \vec{q}_{0},\label{eq:eom_qt_solution_case_III_V}\\
\vec{p}_{t}  &  = \vec{p}_{0} - t \left\langle \partial\mathbf{V}/\partial
\vec{q} (\vec{q}_{0}) \right\rangle _{\boldsymbol{\psi}_{0}(\vec{q}_{0})}.
\label{eq:eom_pt_solution_case_III_V}%
\end{align}

\subsubsection{\label{subsubsec:kin_prop}Kinetic propagation: $\hat{H} =
T(\hat{\vec{p}}\,)$}

Similarly, when $\hat{H}=T(\hat{\vec{p}}\,)$, Eqs.~(\ref{eq:eom_wf_case_I}),
(\ref{eq:eom_qt_case_II}), (\ref{eq:eom_pt_case_II}) become
\begin{align}
\dot{\tilde{\boldsymbol{\psi}}}_{t}(\vec{p}_{t}) &  =-\frac{i}{\hbar}%
\tilde{\mathbf{T}}(\vec{p}_{t})\tilde{\boldsymbol{\psi}}_{t}(\vec{p}%
_{t}),\label{eq:eom_wf_case_III_T}\\
\dot{\vec{q}}_{t} &  =m^{-1}\cdot\langle\tilde{\mathbf{p}}(\vec{p}_{t}%
)\rangle_{\tilde{\boldsymbol{\psi}}_{t}(\vec{p}_{t})}%
,\label{eq:eom_qt_case_III_T}\\
\dot{\vec{p}}_{t} &  =0,\label{eq:eom_pt_case_III_T}%
\end{align}
where $[\tilde{\mathbf{T}}(\vec{p}_{t})]_{KK^{\prime}}=\delta_{KK^{\prime}%
}T\boldsymbol{(}\vec{p}\,^{K}(\vec{p}_{t})\boldsymbol{)}$. The exact solution
of Eqs.~(\ref{eq:eom_wf_case_III_T})--(\ref{eq:eom_pt_case_III_T}) for any
time $t$ is
\begin{align}
\tilde{\boldsymbol{\psi}}_{t}(\vec{p}_{t}) &  =\tilde{\mathbf{U}}_{\text{T}%
}(t,\vec{p}_{0})\tilde{\boldsymbol{\psi}}_{0}(\vec{p}_{0}%
),\label{eq:eom_wf_solution_case_III_T}\\
\vec{q}_{t} &  =\vec{q}_{0}+tm^{-1}\cdot\vec{p}_{0}%
,\label{eq:eom_qt_solution_case_III_T}\\
\vec{p}_{t} &  =\vec{p}_{0}.\label{eq:eom_pt_solution_case_III_T}%
\end{align}
Note that Eqs.~(\ref{eq:eom_pt_solution_case_III_V}) and
(\ref{eq:eom_qt_solution_case_III_T}), which appear to be first-order
approximations, are exact since $\partial\mathbf{V}/\partial\vec{q}(\vec
{q}_{t})$ commutes with $\mathbf{U}_{\text{V}}(t,\vec{q}_{t})$, and
$\tilde{\vec{\mathbf{p}}}(\vec{p}_{t})$ with $\tilde{\mathbf{U}}_{\text{T}%
}(t,\vec{p}_{t})$; therefore, $\langle\partial\mathbf{V}/\partial\vec{q}%
(\vec{q}_{t})\,\rangle_{\boldsymbol{\psi}_{t}(\vec{q}_{t})}$ in
Eq.~(\ref{eq:eom_pt_case_III_V}) and $\langle\tilde{\mathbf{p}}(\vec{p}%
_{t})\rangle_{\tilde{\boldsymbol{\psi}}_{t}(\vec{p}_{t})}$ in
Eq.~(\ref{eq:eom_qt_case_III_T}) are time-independent.
The evaluation of $\mathbf{U}_{\text{V}}(t, \vec{q}_{0})$ in
Eq.~(\ref{eq:eom_wf_solution_case_III_V}) and $\tilde{\mathbf{U}}_{\text{T}}(t, \vec{p}_{0})$ in
Eq.~(\ref{eq:eom_wf_solution_case_III_T}), as well as  their action on $\boldsymbol{\psi}_{0}(\vec{q}_{0})$ and
$\tilde{\boldsymbol{\psi}}_{0}(\vec{p}_{0})$, respectively, can be parallelized using the openMP interface.

The resulting, overall evolution operators (which also include the grid
evolution) for the potential and kinetic splitting steps are
\begin{align}
\mathbf{U}  &  _{\text{V,adpt}}(\Delta t; \boldsymbol{\psi}_{0},\vec{q}%
_{0}):=\mathbf{U}_{\text{V}}(\Delta t, \vec{q}_{0}),\\
\mathbf{U}  &  _{\text{T,adpt}}(\Delta t; \boldsymbol{\psi}_{0},\vec{q}_{0},
\vec{p}_{0})\nonumber\\
&  :=\tilde{\mathbf{f}}(\vec{q}_{\Delta t}, \vec{p}_{0}) \tilde{\mathbf{U}%
}_{\text{T}}(\Delta t, \vec{p}_{0}) \mathbf{f}(\vec{q}_{0}, \vec{p}_{0}),
\label{eq:T_split_substep_adaptive}%
\end{align}
respectively, where we have used that $\vec{p}_{0} = \vec{p}_{\Delta t}$ in
Eq.~(\ref{eq:T_split_substep_adaptive}) [which follows from
Eq.~(\ref{eq:eom_pt_solution_case_III_T})]. Evolution operators $\mathbf{U}%
_{\text{V,adpt}}(\Delta t; \boldsymbol{\psi}_{0},\vec{q}_{0})$ and
$\mathbf{U}_{\text{T,adpt}}(\Delta t; \boldsymbol{\psi}_{0},\vec{q}_{0},
\vec{p}_{0})$ preserve the norm, which follows from
Eq.~(\ref{eq:F_inv_definition}) and from the fact that $\mathbf{U}_{\text{V}%
}(\Delta t, \vec{q}_{0})^{\dagger} \mathbf{U}_{\text{V}}(\Delta t, \vec{q}%
_{0}) = \tilde{\mathbf{U}}_{\text{T}}(\Delta t, \vec{p}_{0})^{\dagger}
\tilde{\mathbf{U}}_{\text{T}}(\Delta t, \vec{p}_{0}) = 1$. A composition of
norm-preserving operators is norm-preserving\cite{Choi_Vanicek:2019} and,
therefore, any split-operator algorithm composed from $\mathbf{U}%
_{\text{V,adpt}}(\Delta t; \boldsymbol{\psi}_{0},\vec{q}_{0})$ and
$\mathbf{U}_{\text{T,adpt}}(\Delta t; \boldsymbol{\psi}_{0},\vec{q}_{0},
\vec{p}_{0})$ is norm-preserving.

Time reversibility of $\mathbf{U}_{\text{V,adpt}}(\Delta t; \boldsymbol{\psi
}_{0},\vec{q}_{0})$ follows because $\mathbf{U}_{\text{V}}(-\Delta t, \vec
{q}_{0}) \mathbf{U}_{\text{V}}(\Delta t, \vec{q}_{0}) = 1$. Similarly,
$\mathbf{U}_{\text{T,adpt}}(\Delta t; \boldsymbol{\psi}_{0},\vec{q}_{0},
\vec{p}_{0})$ is time-reversible because
\begin{align}
\mathbf{U}_{\text{T,adpt}}(-\Delta t  &  ; \boldsymbol{\psi}_{\Delta t}%
,\vec{q}_{\Delta t}, \vec{p}_{0})\mathbf{U}_{\text{T,adpt}}(\Delta t;
\boldsymbol{\psi}_{0},\vec{q}_{0}, \vec{p}_{0})\nonumber\\
&  = \tilde{\mathbf{f}}(\vec{q}_{0}, \vec{p}_{0}) \tilde{\mathbf{U}}%
_{\text{T}}(- \Delta t, \vec{p}_{0}) \mathbf{f}(\vec{q}_{\Delta t}, \vec
{p}_{0})\nonumber\\
&  \times\tilde{\mathbf{f}}(\vec{q}_{\Delta t}, \vec{p}_{0}) \tilde
{\mathbf{U}}_{\text{T}}(\Delta t, \vec{p}_{0}) \mathbf{f}(\vec{q}_{0}, \vec
{p}_{0}) = 1,
\end{align}
where we used Eq.~(\ref{eq:F_inv_definition}) and the identity $\mathbf{U}%
_{\text{T}}(-\Delta t, \vec{p}_{0}) \mathbf{U}_{\text{T}}(\Delta t, \vec
{p}_{0}) = 1$. A symmetric composition of time-reversible operators is
time-reversible.\cite{Choi_Vanicek:2019} Therefore, all symmetric
split-operator algorithms of form~(\ref{eq:comp_TVT}) that are composed from
$\mathbf{U}_{\text{V,adpt}}(\Delta t; \boldsymbol{\psi}_{0},\vec{q}_{0})$ and
$\mathbf{U}_{\text{T,adpt}}(\Delta t; \boldsymbol{\psi}_{0},\vec{q}_{0},
\vec{p}_{0})$ are time-reversible.

\subsection{\label{subsec:stability}Stability of the time-reversible adaptive
grid}

Equations~(\ref{eq:eom_qt_solution_case_III_V}),
(\ref{eq:eom_pt_solution_case_III_V}), (\ref{eq:eom_qt_solution_case_III_T}),
and (\ref{eq:eom_pt_solution_case_III_T}) for the evolution of the grid
centers are essentially the equations of the Verlet
algorithm.\cite{Verlet:1967,book_Leimkuhler_Reich:2004,Frenkel_Smit:2002} The
stability\cite{book_Leimkuhler_Reich:2004, book_Bhatia_George:2006} of the
method from Sec.~\ref{subsec:recovery_of_symmetry_and_time_reversibility},
therefore, depends mostly on the stability of the Verlet algorithm because the
split-operator algorithms, by themselves, are stable for all $\Delta t$.

The Verlet algorithm applied to the harmonic oscillator is stable for time
steps that satisfy
\begin{equation}
\Delta t < T_{\text{osc}} / \pi, \label{eq:verlet_stability_cond}%
\end{equation}
where $T_{\text{osc}}$ is the oscillation
period.\cite{book_Leimkuhler_Reich:2004} In higher-dimensional harmonic
models, the restriction~(\ref{eq:verlet_stability_cond}) on the time step must
hold for the period $T_{\text{osc}}$ of the fastest normal
mode.\cite{book_Leimkuhler_Reich:2004}

\section{\label{sec:numerical_examples}Numerical examples}

\subsection{\label{subsec:harmonic}Three-dimensional harmonic model}

To analyze the geometric and convergence properties of the algorithm proposed
in Sec.~\ref{subsec:recovery_of_symmetry_and_time_reversibility}, we devised a
two-surface three-dimensional harmonic model of electronic excitation of a
molecule. The initial vibrational state, determined using the ground-state
potential energy surface, was propagated solely on the excited-state surface,
following an impulsive electronic excitation. More precisely, the initial
state for the propagation was the ground vibrational eigenstate,
\begin{equation}
\psi(\vec{q}\,) = (\pi\hbar)^{-D/4} \exp(-\vec{q} \, ^{2} / 2 \hbar),
\label{eq:gaussian_wavepacket}%
\end{equation}
of the ground-state Hamiltonian,
\begin{equation}
\hat{H}_{g} = \sum_{l=1}^{D}\frac{\omega_{l}}{2} \left[  (\hat{p}_{l})^{2} +
(\hat{q}_{l})^{2} \right]  , \label{eq:g_ham_scaled}%
\end{equation}
where $q_{l}$ is the $l$th ground-state normal mode coordinate, $p_{l}$ is its
conjugate momentum, and $\omega_{l}$ is the associated vibrational frequency.
After the electronic excitation, $\psi(\vec{q}\,)$ was propagated with the
excited-state Hamiltonian
\begin{equation}
\hat{H}_{e} = \sum_{l=1}^{D} \frac{\omega_{l}}{2} \hat{p}_{l}^{2} + \frac
{1}{2} (\hat{\vec{q}} - \vec{q}_{0})^{T} \cdot K \cdot(\hat{\vec{q}} - \vec
{q}_{0}), \label{eq:e_ham_scaled}%
\end{equation}
where $\vec{q}_{0}$ is the displacement of the excited-state potential energy
surface and $K$ is a symmetric positive definite matrix; $K$ is not diagonal
because the excited-state normal modes were chosen to be Duschinsky
rotated\cite{Duschinsky:1937} with respect to the ground-state normal modes.
For the dynamics, natural units (n.u.) were used: $\hbar= \omega_{2} =
m_{\text{H}} = 1$, where $m_{\text{H}}$ is the mass of a hydrogen atom. The
diagonal ($K_{ll}$) and off-diagonal ($K_{lm}$) elements of the $K$ matrix,
displacement $\vec{q}_{0}$, and ground-state vibrational frequency $\omega
_{l}$ in Eq.~(\ref{eq:e_ham_scaled}) are listed in
Table~\ref{tab:gen_vib_coefficients_init_cond}, which also contains the
initial parameters of the adaptive grid and the total propagation time $t_{f}$.

\begin{table}
[th]%
\caption{Parameters for the quantum dynamics of the harmonic model from Sec.~\ref{subsec:harmonic}. The parameters of the Hamiltonian~(\ref{eq:e_ham_scaled}) (the force constant $K$, displacement $\vec{q}_{0}$, and ground-state vibrational frequencies $\vec{\omega}$), initial parameters of the adaptive grid ($\vec{N}, \vec{q}_{\text{ctr}}, \Delta \vec{q}\,$), and the total propagation time are shown in natural units (n.u.) defined in Sec.~\ref{subsec:harmonic}.}
\label{tab:gen_vib_coefficients_init_cond} \begin{ruledtabular}
\begin{tabular}{crcr}
Parameters & \multicolumn{1}{c}{Values} & Parameters & \multicolumn{1}{c}{Values}\\
\hline
$K_{11}$              & $1.997$                  &  $\omega_{1}$               & $2$\\
$K_{22}$              & $1.015$                  &  $\omega_{2}$                   &$1$ \\
$K_{33}$              & $2.48$                &     $\omega_{3}$                & $2.5$\\
$K_{12}$              & $-0.04$   & $N_{1} = N_{2} = N_{3}$  &  $32$ \\
$K_{13}$              & $-0.017$  &$q_{\text{ctr}, 1}= q_{\text{ctr}, 2} = q_{\text{ctr},3}$   &  $0$ \\
$K_{23}$              & $0.04$    &$\Delta q_{1}= \Delta q_{2} = \Delta q_{3}$            &  $0.4375$ \\
$q_{0,1}$& $-7$                  &$t_{f}$              & $50$ \\
$q_{0,2} = q_{0,3}$ & $7$       & &
\end{tabular}
\end{ruledtabular}

\end{table}

To verify that grid adaptation does not decrease the accuracy of the solution,
we compared the wavefunction $\psi_{t}^{(\Delta t)}$ propagated using the
adaptive grid with the time step $\Delta t$ to the corresponding ``benchmark''
wavefunction $\Psi_{t}^{(\Delta t)}$ propagated using a fixed grid. Indeed,
the errors $\| \psi_{t}^{(\Delta t)} - \Psi_{t}^{(\Delta t)} \|$ were
minuscule (the errors were $5 \times10^{-11}$ at $t=0$, and $2 \times10^{-10}$
at $t=t_{f}$). The wavefunctions were propagated with the optimally composed
tenth-order TVT split-operator algorithm with $\Delta t = t_{f}/2^{9}$. (See
Ref.~\onlinecite{Choi_Vanicek:2019} and the references therein for a detailed
discussion of composition schemes.) We used a high-order integrator with a
small time step so that the error was dominated by grid adaptation, and not by
time discretization. Both $\vec{q}$- and $\vec{p}$-ranges of the fixed grid
were chosen to be twice larger than the ranges of the adaptive grid because
the amplitude of the adaptive grid's oscillation was approximately equal to
its range. In order that the fixed and adaptive grids had the same density,
the fixed grid was chosen to have $128 \times128 \times128$ points (see Appendix~\ref{appendix:grid_conv} for the exponential convergence of the wavefunction with the number of grid points).

\begin{figure}
[tbp]\includegraphics[]{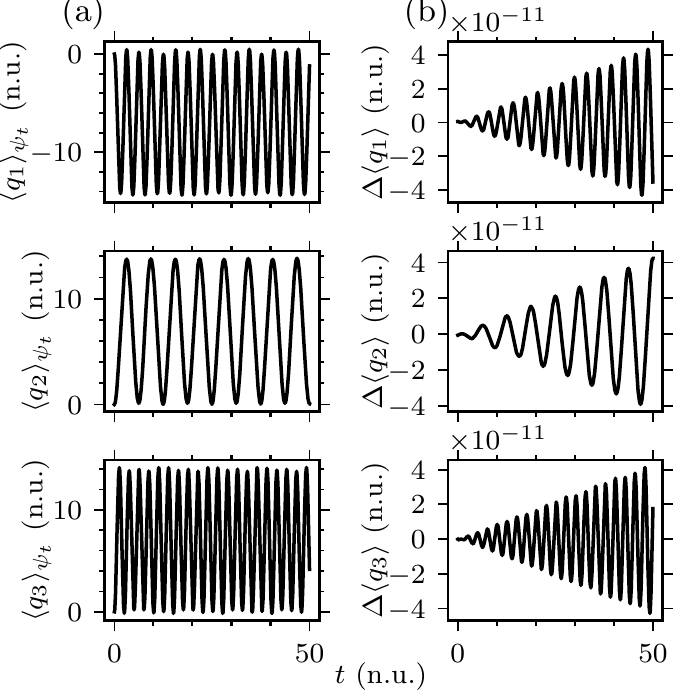} \caption{Accuracy of the adaptive grid used for quantum dynamics of the three-dimensional harmonic model~(\ref{eq:e_ham_scaled}) from Sec.~\ref{subsec:harmonic}. (a) Position expectation values $\langle q_{l} \rangle_{\psi_{t}}$ computed on the adaptive grid. (b) Difference $\Delta \langle q_{l} \rangle := \langle q_{l} \rangle_{\psi_{t}} - \langle q_{l} \rangle_{\Psi_{t}}$ between the position expectation values computed on the adaptive grid ($32 \times 32 \times 32$ points) and fixed grid ($128 \times 128 \times 128$ points). \label{fig:harm_q_vs_t}}
\end{figure}

Figure~\ref{fig:harm_q_vs_t}(a) shows that the expectation value of position
is computed correctly with the adaptive grid, even when the wavefunction moves
beyond the range of the initial grid. In fact, Fig.~\ref{fig:harm_q_vs_t}(b)
shows that the error of the position expectation value is minuscule (of the
order of $10^{-11}$) for all times; the slow linear increase in the error is
due to the accumulation of roundoff errors.

\begin{figure*}
[ptb]%
\includegraphics[]{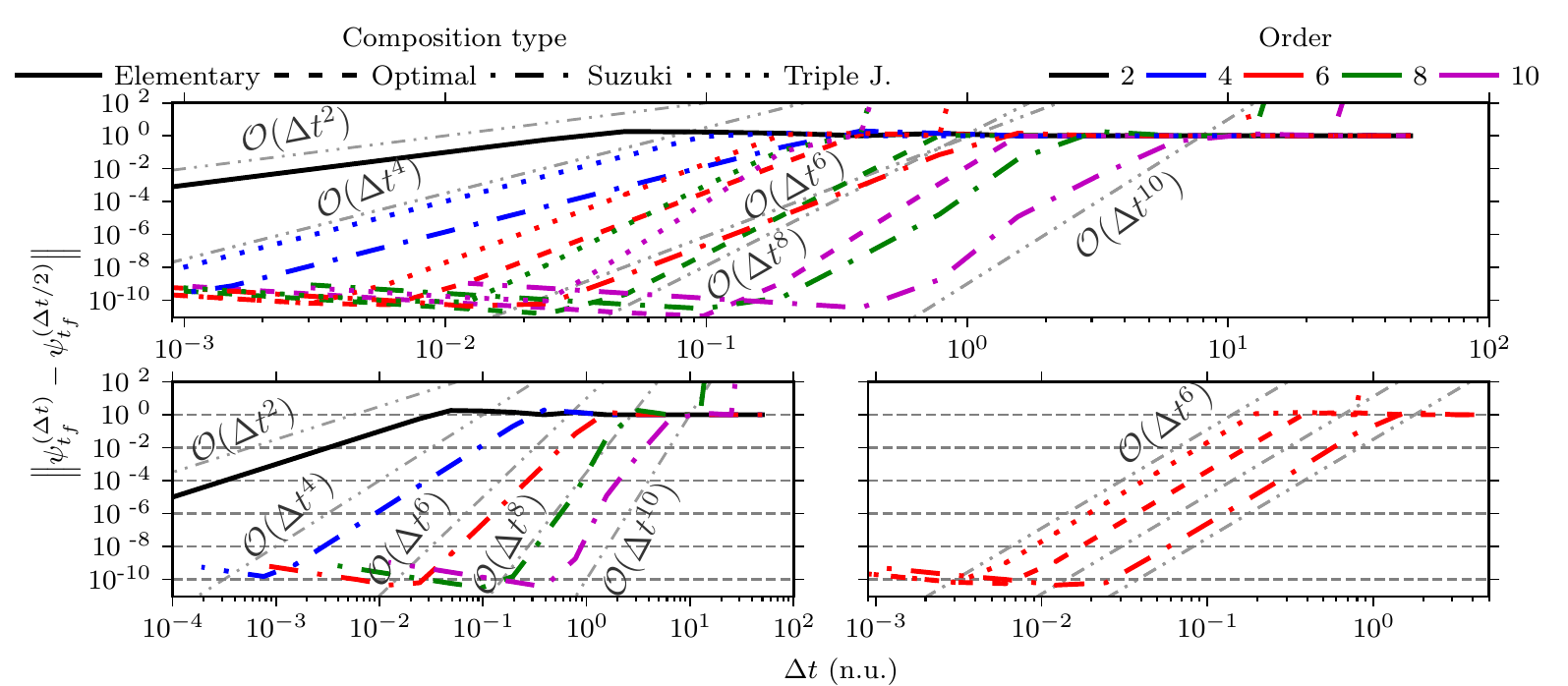}\caption{Convergence (up to the
tenth-order) of the wavefunction as a function of the time step in the
harmonic system from Sec.~\ref{subsec:harmonic}. In this figure and
Figs.~\ref{fig:harm_efficiency}--\ref{fig:harm_t_rev_fail}, we only show the
results for the compositions of the TVT algorithm. Gray straight lines
indicate predicted orders of convergence. Top: all discussed methods; bottom
left: methods composed through Suzuki's fractal\cite{Suzuki:1990}; bottom
right: sixth-order methods.}\label{fig:harm_convergence}
\end{figure*}

Figure~\ref{fig:harm_convergence} demonstrates that the
compositions\cite{book_Hairer_Wanner:2006, Suzuki:1990, Yoshida:1990,
Kahan_Li:1997, Sofroniou_Spaletta:2005} of the proposed algorithm from
Sec.~\ref{subsec:recovery_of_symmetry_and_time_reversibility} achieve the
predicted higher orders of accuracy. The figure also demonstrates the
divergence of the discretization errors $\| \psi^{(\Delta t)}_{t_{f}} -
\psi^{(\Delta t/2)}_{t_{f}} \|$ when the composition substep size,
$|\gamma_{k}| \Delta t$, does not satisfy
condition~(\ref{eq:verlet_stability_cond}). Note that the accumulation of
roundoff errors does not allow the discretization errors to reach below
$\approx10^{-10}$.

\begin{figure}
[tbp]\includegraphics[]{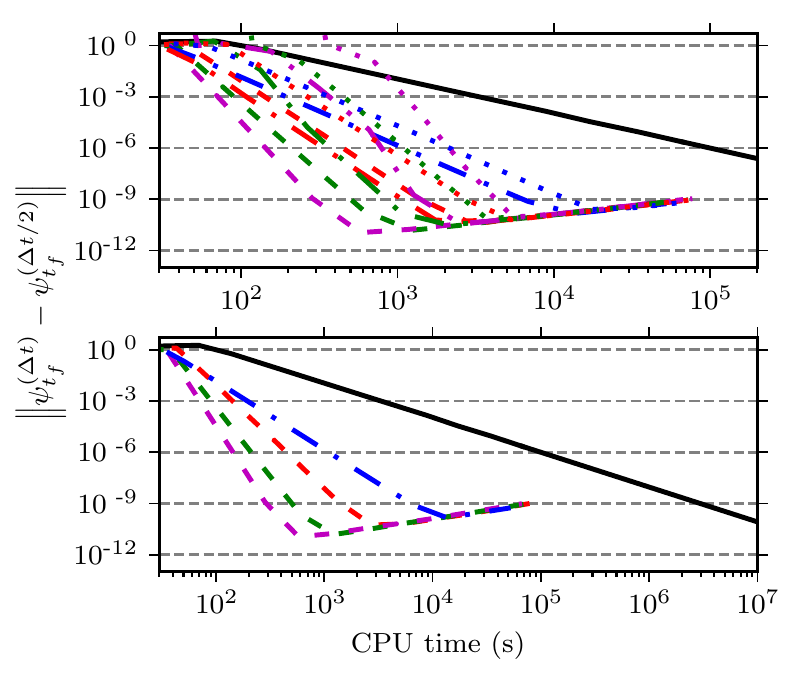} \caption{Efficiency of the compositions of the proposed algorithm (see Sec.~\ref{subsec:recovery_of_symmetry_and_time_reversibility}) in the harmonic system from Sec.~\ref{subsec:harmonic}. Results of the second-order TVT algorithm were extrapolated using the line of best fit beyond CPU time $= 7 \times 10^{4}$ s to highlight the higher efficiency of the higher-order methods. Top: all discussed methods; bottom: optimally composed methods (Suzuki's fractal is the optimal fourth-order composition scheme\cite{Kahan_Li:1997, Sofroniou_Spaletta:2005, Choi_Vanicek:2019}). Line labels are the same as in Fig.~\ref{fig:harm_convergence}. \label{fig:harm_efficiency}}
\end{figure}

Higher-order methods require a longer computational time per time step, but converge faster with the
decreasing time step size. Therefore, to obtain a wavefunction with a discretization error below a certain threshold
value, higher-order methods have a lower computational cost, which we measure by the central processing unit
(CPU) time. Figure~\ref{fig:harm_efficiency} shows that a tenfold speedup is
already achieved by using the optimal tenth-order instead of the second-order
algorithm to reach a moderate discretization error of $10^{-2}$.
The speedup relative to the second-order algorithm is much greater if a small error is desired.
To reach an error of $10^{-9}$, $500$-fold speedup is achieved by using the Suzuki fourth-order algorithm.
Moreover, $2000$-, $5000$-, and $10000$-fold speedups are achieved by using the optimal sixth-, eighth-, and
tenth-order algorithms, respectively.

\begin{figure}
[tbp]\includegraphics[]{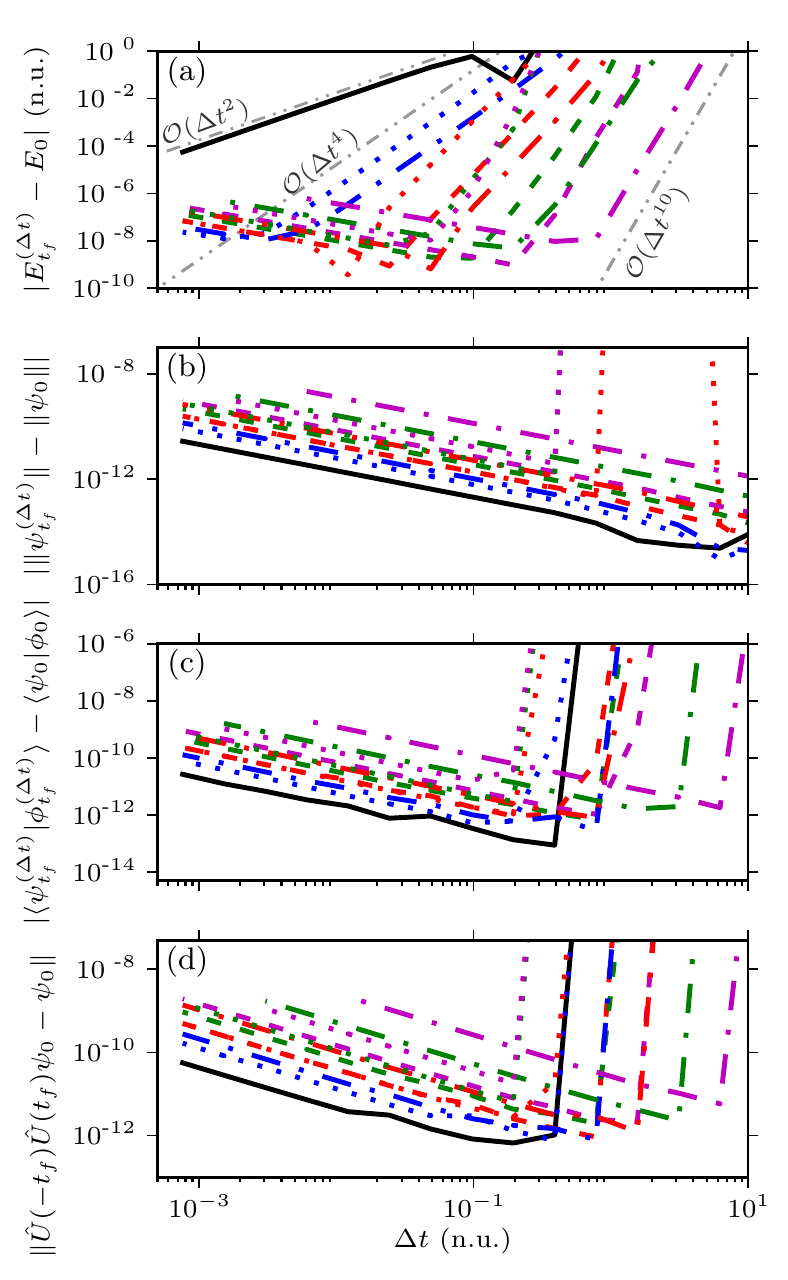} \caption{Conservation of geometric properties by the compositions of the proposed algorithm from Sec.~\ref{subsec:recovery_of_symmetry_and_time_reversibility} as a function of the time step in the harmonic system from Sec.~\ref{subsec:harmonic}: (a) energy [$E_{0} = 140.389$ n.u.], (b) norm, (c) inner product, and (d) time reversibility. $\phi_{0}$ is wavepacket~(\ref{eq:gaussian_wavepacket}) displaced by $q_{1} = -1$ n.u. and $q_{2} = q_{3} = 1$ n.u. (hence, $\langle \psi_{0} | \phi_{0} \rangle  = 0.472$). Time reversibility is measured by the distance between $\psi_{0}$ and the forward-backward propagated state, i.e., $\psi_{0}$ propagated forward in time for $t_{f}$ and then backward in time for $t_{f}$. Gray straight lines indicate predicted orders of convergence $\mathcal{O}(\Delta t^{m})$ for $m=2,4,$ and $10$. Line labels are the same as in Fig.~\ref{fig:harm_convergence}. \label{fig:harm_geom_prop}}
\end{figure}

Because a split-operator propagation is equivalent to an exact propagation
with an effective, time-dependent Hamiltonian, the energy $E_{t}^{(\Delta t)}
= \langle\psi_{t}^{(\Delta t)} | \hat{H} | \psi_{t}^{(\Delta t)} \rangle$ is
conserved only approximately. Figure~\ref{fig:harm_geom_prop}(a) shows that
the energy is conserved to the same order of accuracy [$\mathcal{O}(\Delta
t^{m})$] as the wavefunction. Figures~\ref{fig:harm_geom_prop}(b) and (d)
demonstrate that the compositions of the proposed algorithm are exactly
norm-preserving and time-reversible as already justified analytically in
Sec.~\ref{subsec:recovery_of_symmetry_and_time_reversibility}. In
Sec.~\ref{subsec:loss_of_geom_prop}, we showed that the grid adaptation leads
to the non-conservation of the inner product. Figure~\ref{fig:harm_geom_prop}%
(c) may, therefore, be misleading because the inner product appears to be
conserved. However, this is not true in general, as shown later on the example
of collinear He--H$_{2}$ scattering (see Sec.~\ref{subsec:scattering}).

In all panels of Fig.~\ref{fig:harm_geom_prop} the slow increase in the error
for decreasing time steps is due to the accumulation of roundoff errors;
therefore, the (minuscule) errors are larger for methods with more composition
steps per time step.\cite{book_Hairer_Wanner:2006} Panels~(b), (c), and (d)
show that, on the other hand, the errors diverge for large time steps $\Delta
t$ because of the instability of the Verlet algorithm (see
Sec.~\ref{subsec:stability}); larger errors result from methods with a larger
maximum composition coefficient [$\max_{k}|\gamma_{k}|$, see
Eq.~(\ref{eq:verlet_stability_cond})].
Specifically, for large $\Delta t$, the time reversibility of the proposed algorithm is lost because the centers of the
position and momentum grids diverge due to the instability of the Verlet algorithm, used for propagating
the grid centers. Similarly, beyond a certain time step size, the norm is no longer preserved because it is evaluated
on a grid whose center has diverged to infinity.

\begin{figure}
[tbp]\includegraphics[]{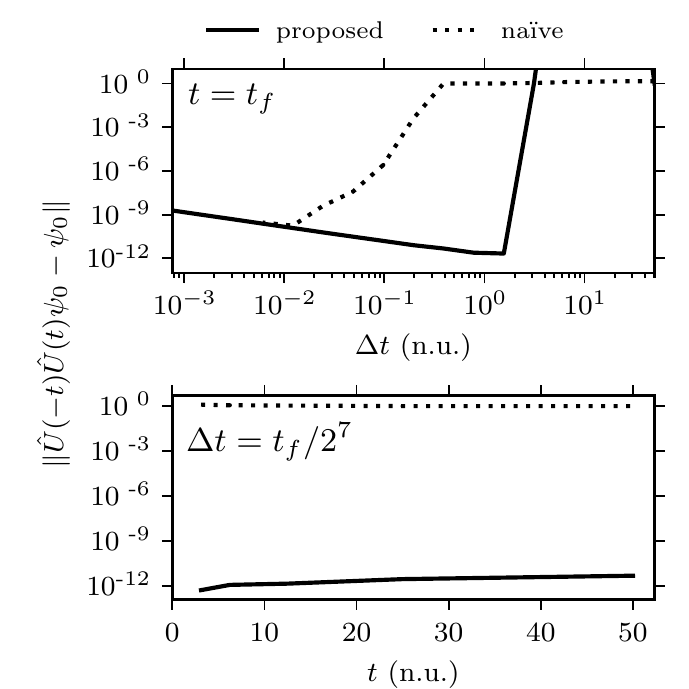} \caption{Time reversibility of the na\"{i}ve (see Sec.~\ref{subsec:loss_of_geom_prop_naive}) and proposed (see Sec.~\ref{subsec:recovery_of_symmetry_and_time_reversibility}) split-operator algorithms on adaptive grids in the harmonic system from Sec.~\ref{subsec:harmonic}. Both split-operator algorithms were composed to the tenth order using the optimal scheme.  Top: time reversibility as a function of the time step $\Delta t$ for a fixed total propagation time $t$. Bottom: time reversibility as a function of the total propagation time $t$ for a fixed time step $\Delta t$. \label{fig:harm_t_rev_fail}}
\end{figure}

Figure~\ref{fig:harm_t_rev_fail} confirms that the na\"{\i}ve adaptive grid
approach is not time-reversible while the algorithm proposed in
Sec.~\ref{subsec:recovery_of_symmetry_and_time_reversibility} is. Note that
for very small time steps ($\Delta t \leq10^{-2}$), the solution is
essentially exact, and even the na\"{\i}ve adaptive grid approach becomes
effectively time-reversible. The bottom panel of
Fig.~\ref{fig:harm_t_rev_fail} shows, however, that for a fixed time step
$\Delta t$, the time propagation on the na\"{\i}ve adaptive grid is not
time-reversible already after a short propagation time $t$; the breaking of
time reversibility is an inherent property of the na\"{\i}ve adaptive grid.

\subsection{\label{subsec:scattering}Collinear He--H$_{2}$ scattering}

As a more challenging test, we also applied the algorithm proposed in
Sec.~\ref{subsec:recovery_of_symmetry_and_time_reversibility} to a very
anharmonic system. Following Ref.~\onlinecite{Sim_Makri:1995}, we simulated
the collinear He--H$_{2}$ scattering using a
modified\cite{Campos-Martinez_Coalson:1990}
Secrest--Johnson\cite{Secrest_Johnson:1966} potential energy surface,
\begin{equation}
V_{\text{SJ}}(\vec{q}) = D(1-e^{-\beta q_{1}})^{2} + e^{-\alpha(q_{2} -
q_{1})}, \label{eq:modified_secrest_johnson}%
\end{equation}
where $\beta=0.158$ n.u., $D=20$ n.u., and $\alpha=0.3$ n.u. The natural units
(n.u.) are different from those defined in Sec.~\ref{subsec:harmonic}: $\hbar=
1$ as before, but $\sqrt{2 D \beta^{2}/m_{1}} = m_{1} = 1$ instead. In
Eq.~(\ref{eq:modified_secrest_johnson}), $q_{1}$ is the vibrational coordinate
of H$_{2}$, and $q_{2}$ is the distance between the He atom and the center of
mass of H$_{2}$.\cite{Secrest_Johnson:1966} In this coordinate system, $m_{1}
= 1$ n.u. and $m_{2} = 2/3$ n.u.\cite{Sim_Makri:1995, Secrest_Johnson:1966}
The Hamiltonian for this problem is $\hat{H}_{\text{scat}} = T(\hat{\vec{p}})
+ V_{\text{SJ}}(\hat{\vec{q}})$, where $T(\hat{\vec{p}})$ has the
form~(\ref{eq:KE_operator}).

Identically to Ref.~\onlinecite{Sim_Makri:1995}, the initial state is a
product of two one-dimensional Gaussian wavepackets
\begin{align}
\psi_{(1)}(q)  &  = (\pi\hbar)^{-1/4} \exp(-q^{2} / 2 \hbar),\nonumber\\
\psi_{(2)}(q)  &  = (\pi\sigma_{0}^{2})^{-1/4} \exp[- (q - q_{0})^{2} / 2
\sigma_{0}^{2} + i p_{0} (q - q_{0})/ \hbar]. \label{eq:general_GWP}%
\end{align}
The Gaussian wavepacket $\psi_{(2)}(q)$ is sufficiently narrow and far from
the interaction region so that there is no significant initial interaction
between He and H$_{2}$ ($\sigma_{0}^{2} = 8$ n.u. and $q_{0} = 24$ n.u.). A
negative initial momentum ($p_{0} = -3.56$ n.u.) ensures a collision at a
later time.

\begin{figure}
[btp]\includegraphics[]{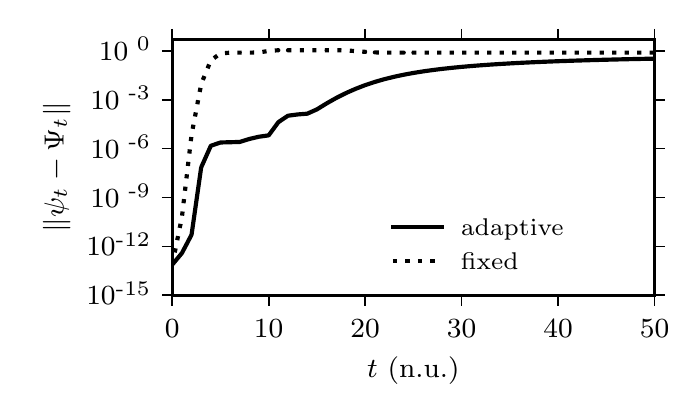}
\caption{Error $\| \psi_{t} - \Psi_{t} \|$ of the wavefunction $\psi_{t}$ propagated on either the adaptive or fixed
grid (both with $128 \times 128$ points) in the He--H$_{2}$ scattering from Sec.~\ref{subsec:scattering}.
$\Psi_{t}$ is the ``exact'' reference wavefunction propagated on a fixed grid with $128 \times 2048$ points.
The initial $q_{1}$ range of all grids was ($-14$ n.u., $14$ n.u.). The initial $q_{2}$ range was
($0$ n.u., $48$ n.u.) for both grids with $128 \times 128$ points and ($-50$ n.u., $400$ n.u.) for the reference
fixed grid. Figures~\ref{fig:scattering_error}--\ref{fig:scattering_geom_prop} were produced using the optimal
tenth-order composition of the VTV algorithm with $\Delta t = 0.1$ n.u. \label{fig:scattering_error}}
\end{figure}

Figure~\ref{fig:scattering_error} shows the error of the wavefunction
propagated using either the adaptive or fixed grid, both with the same number
of grid points. The error of the wavefunction propagated using the adaptive
grid remains reasonably small ($< 10^{-3}$) for six times longer than the
error of the corresponding wavefunction on the fixed grid. The significance is
that for a given number of grid points, determined, e.g., by the available
memory, the time scale of a simulation can be extended by grid adaptation.

\begin{figure}
[tbp]\includegraphics[]{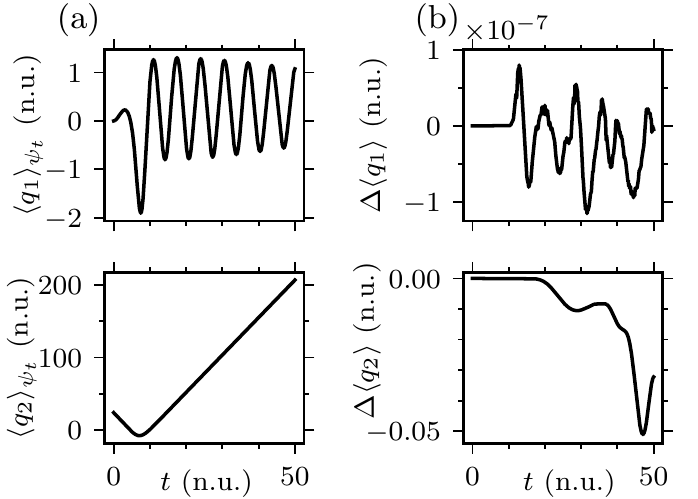} \caption{Accuracy of the adaptive grid for simulating the He--H$_{2}$ scattering (see Sec.~\ref{subsec:scattering}) (a) Expectation values of position computed on the adaptive grid. (b) The difference between the position expectation values computed on the adaptive grid ($ 128 \times 128$ points) and reference fixed grid ($ 128 \times 2048$ points). \label{fig:scattering_q_vs_t}}
\end{figure}

Figure~\ref{fig:scattering_q_vs_t} displays the time dependence of the
expectation value of $\vec{q}$ and of its error. Panel~(a) shows that the
collision between He and H$_{2}$ induces the vibration of H$_{2}$, which was
originally in its ground vibrational state. Panel~(b) shows that the error of
the position expectation value was reasonably small until $t\approx20$ n.u. on
the adaptive grid.
After $t \approx 20$ n.u., however, the error starts to grow rapidly in the second dimension
because the width of the wavepacket in this dimension increases approximately linearly from $t \approx 10$ n.u.
Thus, a significant portion of the wavepacket eventually escapes through the boundaries of the adaptive grid
in the second dimension.

\begin{figure}
[tbp]\includegraphics[]{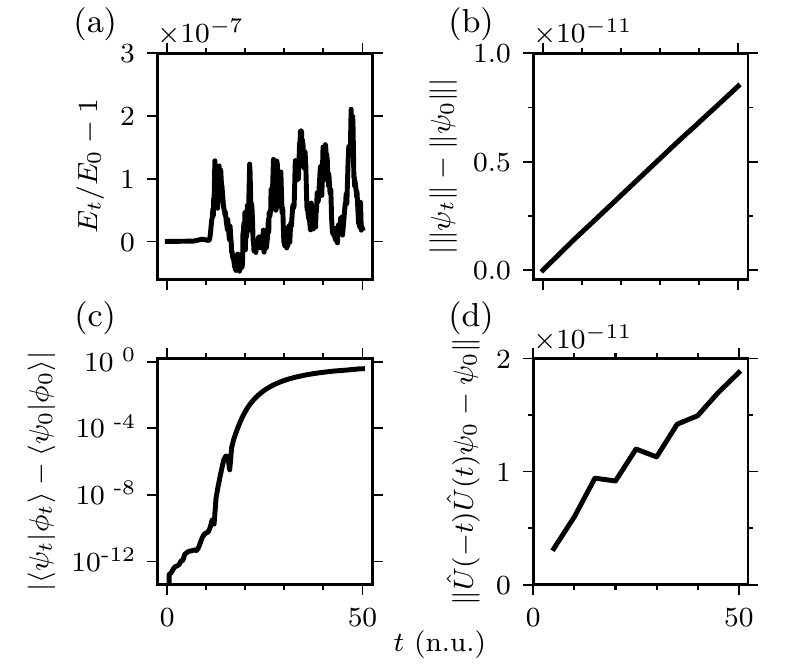} \caption{Geometric properties of the algorithm proposed in Sec.~\ref{subsec:recovery_of_symmetry_and_time_reversibility} applied to the He--H$_{2}$ scattering from Sec.~\ref{subsec:scattering}: (a) energy, (b) norm, (c) inner product, (d) time reversibility. The time reversibility is defined in the same way as in the caption of Fig.~\ref{fig:harm_geom_prop}. Gaussian wavepacket $\phi_{0}$ is identical to $\psi_{0}$, the two-dimensional initial Gaussian wavepacket from Sec.~\ref{subsec:scattering}, except that $p_{0}$ in Eq.~(\ref{eq:general_GWP}) is $-3.0$ n.u. instead of $-3.56$ n.u. (hence, $| \langle \psi_{0} | \phi_{0} \rangle |= 0.534$). \label{fig:scattering_geom_prop}}
\end{figure}

Finally, in Fig.~\ref{fig:scattering_geom_prop}, we show that the proposed
algorithm preserves the geometric invariants even in the collinear scattering
of He--H$_{2}$, where the wavepacket is more delocalized than in the harmonic
example from Sec.~\ref{subsec:harmonic}. As expected, the norm
[Fig.~\ref{fig:scattering_geom_prop}(b)] and time reversibility
[Fig.~\ref{fig:scattering_geom_prop}(d)] are preserved exactly (the slow
linear increase of the invariants is again due to the accumulation of roundoff
errors). On the other hand, the energy [Fig.~\ref{fig:scattering_geom_prop}%
(a)] and inner product [Fig.~\ref{fig:scattering_geom_prop}(c)] are not
conserved. In particular, the apparent conservation of the inner product
observed in Sec.~\ref{subsec:harmonic} is indeed not general.

\subsection{\label{subsec:henonheiles}Eight-dimensional H\'{e}non--Heiles model}

To demonstrate the applicability of the algorithm from Sec.~\ref{subsec:recovery_of_symmetry_and_time_reversibility} to high-dimensional quantum dynamics, we
applied it
to the
eight-dimensional H\'{e}non--Heiles model,
\begin{equation}
\hat{H}_{\text{HH}}=T(\hat{\vec{p}})+V_{\text{HH}}(\hat{\vec{q}}),
\end{equation}
whose kinetic energy $T(\hat{\vec{p}})$ is of the form~(\ref{eq:KE_operator})
and potential energy is given by
\begin{equation}
V_{\text{HH}}(\vec{q})=\frac{\kappa}{2}\sum_{l=1}^{D}q_{l}^{2}+\lambda
\sum_{l=1}^{D-1}(q_{l}^{2}q_{l+1}-q_{l+1}^{3}/3)
\end{equation}
with $D=8$. Simulating the quantum dynamics of the H\'{e}non--Heiles system is
challenging because the potential $V_{\text{HH}}(\vec{q})$ is anharmonic,
unbound, and contains inter-mode couplings. Following
Ref.~\onlinecite{Nest_Meyer:2002}, the parameters were chosen to be
$\lambda=0.111803$ n.u. and $m_{l}=1$ n.u. for $l=1,\dots,D$, where the
natural units (n.u.) were defined by fixing $\hbar = m_{\mathrm{H}} = \kappa = 1$; the initial state is $\psi(\vec{q})=\pi^{-D/4}\exp[-(\vec
{q}-\vec{q}_{0})^{2}/2]$ with $q_{0,l}=2$ n.u. for $l=1,\dots,D$.

\begin{figure}
[tbp]\includegraphics[]{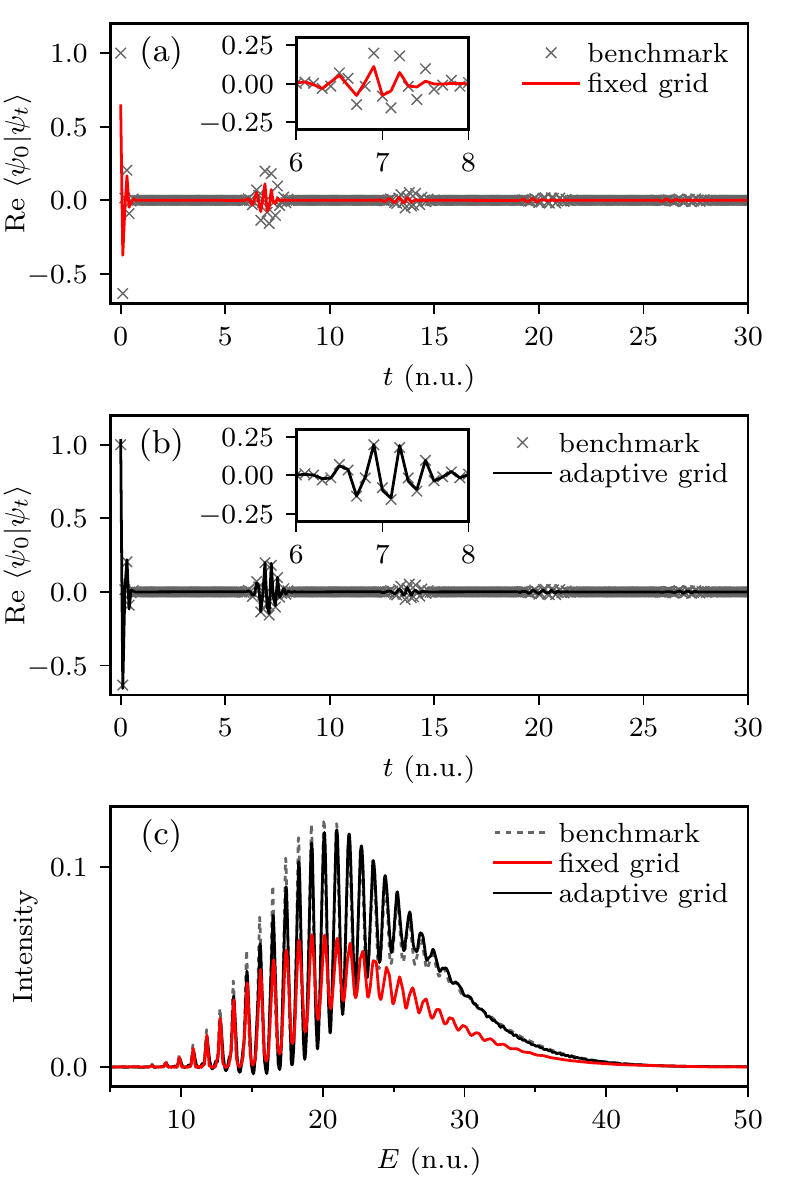} \caption{Autocorrelation function for the H\'{e}non--Heiles model from Sec.~\ref{subsec:henonheiles} computed using either (a) the split-operator algorithm on a fixed grid or (b) the proposed adaptive grid algorithm (see Sec.~\ref{subsec:recovery_of_symmetry_and_time_reversibility}). Both algorithms were composed to the fourth order using Suzuki's fractal. (c) Spectrum calculated via the Fourier transform of the autocorrelation function, using $f(t) = \exp[-(t/t_{\mathrm{damp}})^{2}]$ with $t_{\mathrm{damp}} = 30$ n.u. as the damping function. The fixed grid was defined between $q_{l} = -3.5$ n.u. and $q_{l} = 3.5$ n.u., and the initial adaptive grid between $q_{l} = - 3.0$ n.u. and $q_{l} = 7.0$ n.u.; both grids had $8^{8}$ points. The benchmark was calculated using MCTDH.\cite{Meyer_Cederbaum:1990} \label{fig:henonheiles}}
\end{figure}

Panels~(a) and (b) of Fig.~\ref{fig:henonheiles} show that the autocorrelation
function $\langle\psi_{0}|\psi_{t}\rangle$ is obtained more accurately by the
adaptive grid algorithm
than by the
standard split-operator algorithm on a fixed grid when both the adaptive and
fixed grids have the same number ($8^{8}$) of grid points. Especially, the
detailed shape of the first recurrence of the autocorrelation function is
correctly described only by the proposed algorithm [see insets of panels~(a)
and (b) of Fig.~\ref{fig:henonheiles}]. Consequently, only the spectrum
calculated using the adaptive grid algorithm has the correct shape of
the envelope [see Fig.~\ref{fig:henonheiles}(c)].

\section{conclusion\label{sec:conclusion}}

We have described a split-operator algorithm combined with an adaptive
phase space grid whose center moves according to the wavepacket's expectation
values of position and momentum. By propagating the grid center exactly and
simultaneously with the wavefunction, the symmetry and time reversibility were
built into the proposed algorithm. Adapting the grid reduces the number of
required grid points while maintaining high accuracy in situations where the
wavepacket remains localized. Examples include harmonic systems or short-time
dynamics in moderately anharmonic systems. On the example of He--H$_{2}$
scattering, we showed that the proposed algorithm is also suitable for
longer-time dynamics if only a moderate accuracy of the wavepacket is
required, i.e., when one can ignore small parts of the wavepacket escaping
through the boundaries of the adaptive grid. The algorithm allowed us to compute accurately the medium-resolution spectrum of the eight-dimensional H\'{e}non--Heiles system, which is not only high-dimensional but also highly nonlinear. 

We showed both analytically and numerically that the time reversibility is
lost by the na\"{\i}ve grid adaptation. Then, we introduced an amendment that
recovered the time reversibility. The geometric properties of the resulting
algorithm, namely norm preservation, conditional stability, symmetry, and time
reversibility, were demonstrated analytically as well as numerically on two
different model systems. Note that because neither the Chebyshev\cite{Tal-Ezer_Kosloff:1984} nor the Lanczos\cite{Lanczos:1950,Park_Light:1986} method is time reversible,\cite{book_Lubich:2008} it is only reasonable to combine them with the na\"{i}ve adaptive grid. Moreover, combining the Chebyshev method with adaptive grids would be inappropriate because the advantage of the Chebyshev algorithm as a global, long time propagator would be lost by changing the grid and, therefore, the Hamiltonian matrix regularly at small time intervals.

Because of its symmetry, the proposed algorithm can be composed to obtain
higher-order integrators. We verified that these higher-order integrators are
more efficient compared to the second-order integrator if high accuracy is
desired. As an additional benefit, the proposed algorithm requires no
adjustable parameters for the grid adaptation because the grid center follows
the exact trajectory of the wavepacket's expectation values of position and momentum.

Finally, we hope that the proposed time-reversible integrator for the
time-dependent Schr\"{o}dinger equation on an adaptive grid could serve as a
benchmark for more approximate methods, such as the thawed Gaussian
approximation,\cite{Heller:1975, Wehrle_Vanicek:2014, Begusic_Vanicek:2019}
that rely on the wavepacket remaining localized for relevant time scales.

The authors acknowledge the financial support from the European Research
Council (ERC) under the European Union's Horizon 2020 research and innovation
programme (grant agreement No. 683069 -- MOLEQULE) and thank Tomislav
Begu\v{s}i\'{c} and Lipeng Chen for useful discussions.

\appendix

\section{\label{appendix:grid_conv} Exponential convergence with the number of
grid points}

\begin{figure}
[hbp]\includegraphics[]{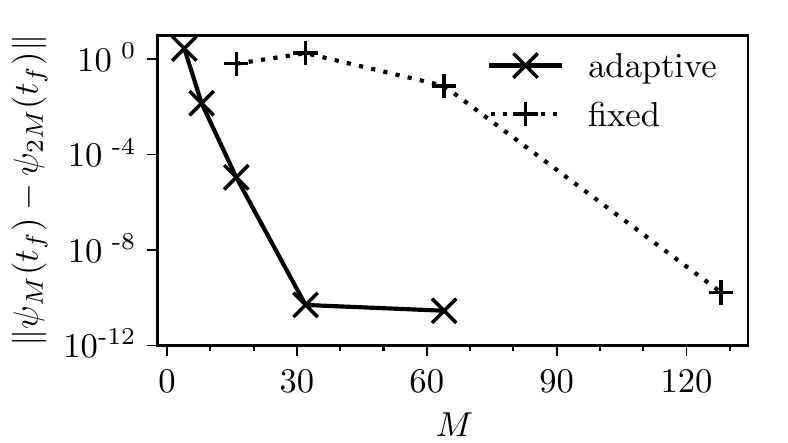} \caption{Convergence of the wavefunction $\psi_{M}(t_{f})$ with the increasing number of grid points. $\psi_{M}(t_{f})$ is evaluated on either the adaptive or fixed grid of $N = M^{D}$ points in the harmonic system from Sec.~\ref{subsec:harmonic}. For balanced $\vec{q}$- and $\vec{p}$-grids, the ranges as well as the densities of both $\vec{q}$- and $\vec{p}$-grids were increased by a factor of $\sqrt{2}$ for every doubling of the number of grid points. Wavefunctions on grids of different densities were compared through the trigonometric interpolation of the wavefunction on the sparser grid. Both the proposed adaptive grid algorithm from Sec.~\ref{subsec:recovery_of_symmetry_and_time_reversibility} and the split-operator algorithm on a fixed grid were composed to the tenth order using the optimal scheme; $\Delta t = 0.1$ n.u.\label{fig:grid_conv}}
\end{figure}

Figure~\ref{fig:grid_conv} shows the exponential convergence of the
wavefunction with the increasing number of grid points. As expected, the
convergence is slower for the fixed grid, which must cover roughly four
times larger phase space area in each dimension to account for the movement of
the wavepacket.

The fast Fourier transform, which is the bottleneck of the combination of the
dynamic Fourier method with the split-operator algorithm on a tensor-product
grid, scales as $N\log N$. Because $N=M^{D}$, where $M$ is the geometric
average number of grid points in each dimension, the method scales
exponentially with the number of dimensions. Similarly to the MCTDH or
time-dependent DVR methods, the proposed algorithm does not remove the
exponential scaling, but slows down the exponential growth significantly
compared to the conventional split-operator algorithm on a fixed grid (see,
e.g., Fig.~\ref{fig:grid_conv}). This is, indeed, what allowed us to treat the
8-dimensional H\'{e}non-Heiles system in Sec.~\ref{subsec:henonheiles}.

\bibliographystyle{aipnum4-1}
\bibliography{moving_grids}

\begin{thebibliography}{73}%
\makeatletter
\providecommand \@ifxundefined [1]{%
 \@ifx{#1\undefined}
}%
\providecommand \@ifnum [1]{%
 \ifnum #1\expandafter \@firstoftwo
 \else \expandafter \@secondoftwo
 \fi
}%
\providecommand \@ifx [1]{%
 \ifx #1\expandafter \@firstoftwo
 \else \expandafter \@secondoftwo
 \fi
}%
\providecommand \natexlab [1]{#1}%
\providecommand \enquote  [1]{``#1''}%
\providecommand \bibnamefont  [1]{#1}%
\providecommand \bibfnamefont [1]{#1}%
\providecommand \citenamefont [1]{#1}%
\providecommand \href@noop [0]{\@secondoftwo}%
\providecommand \href [0]{\begingroup \@sanitize@url \@href}%
\providecommand \@href[1]{\@@startlink{#1}\@@href}%
\providecommand \@@href[1]{\endgroup#1\@@endlink}%
\providecommand \@sanitize@url [0]{\catcode `\\12\catcode `\$12\catcode
  `\&12\catcode `\#12\catcode `\^12\catcode `\_12\catcode `\%12\relax}%
\providecommand \@@startlink[1]{}%
\providecommand \@@endlink[0]{}%
\providecommand \url  [0]{\begingroup\@sanitize@url \@url }%
\providecommand \@url [1]{\endgroup\@href {#1}{\urlprefix }}%
\providecommand \urlprefix  [0]{URL }%
\providecommand \Eprint [0]{\href }%
\providecommand \doibase [0]{http://dx.doi.org/}%
\providecommand \selectlanguage [0]{\@gobble}%
\providecommand \bibinfo  [0]{\@secondoftwo}%
\providecommand \bibfield  [0]{\@secondoftwo}%
\providecommand \translation [1]{[#1]}%
\providecommand \BibitemOpen [0]{}%
\providecommand \bibitemStop [0]{}%
\providecommand \bibitemNoStop [0]{.\EOS\space}%
\providecommand \EOS [0]{\spacefactor3000\relax}%
\providecommand \BibitemShut  [1]{\csname bibitem#1\endcsname}%
\let\auto@bib@innerbib\@empty
\bibitem [{\citenamefont {Heller}(2018)}]{book_Heller:2018}%
  \BibitemOpen
  \bibfield  {author} {\bibinfo {author} {\bibfnamefont {E.~J.}\ \bibnamefont
  {Heller}},\ }\href@noop {} {\emph {\bibinfo {title} {The semiclassical way to
  dynamics and spectroscopy}}}\ (\bibinfo  {publisher} {Princeton University
  Press},\ \bibinfo {address} {Princeton, NJ},\ \bibinfo {year}
  {2018})\BibitemShut {NoStop}%
\bibitem [{\citenamefont {Engel}\ \emph {et~al.}(1988)\citenamefont {Engel},
  \citenamefont {Metiu}, \citenamefont {Almeida}, \citenamefont {Marcus},\ and\
  \citenamefont {Zewail}}]{Engel_Zewail:1988}%
  \BibitemOpen
  \bibfield  {author} {\bibinfo {author} {\bibfnamefont {V.}~\bibnamefont
  {Engel}}, \bibinfo {author} {\bibfnamefont {H.}~\bibnamefont {Metiu}},
  \bibinfo {author} {\bibfnamefont {R.}~\bibnamefont {Almeida}}, \bibinfo
  {author} {\bibfnamefont {R.}~\bibnamefont {Marcus}}, \ and\ \bibinfo {author}
  {\bibfnamefont {A.~H.}\ \bibnamefont {Zewail}},\ }\href {\doibase
  10.1016/0009-2614(88)87319-6} {\bibfield  {journal} {\bibinfo  {journal}
  {Chem.\ Phys.\ Lett.}\ }\textbf {\bibinfo {volume} {152}},\ \bibinfo {pages}
  {1} (\bibinfo {year} {1988})}\BibitemShut {NoStop}%
\bibitem [{\citenamefont {Kosloff}(1988)}]{Kosloff:1988}%
  \BibitemOpen
  \bibfield  {author} {\bibinfo {author} {\bibfnamefont {R.}~\bibnamefont
  {Kosloff}},\ }\href {\doibase 10.1021/j100319a003} {\bibfield  {journal}
  {\bibinfo  {journal} {J.~Phys.\ Chem.}\ }\textbf {\bibinfo {volume} {92}},\
  \bibinfo {pages} {2087} (\bibinfo {year} {1988})}\BibitemShut {NoStop}%
\bibitem [{\citenamefont {Stock}\ \emph {et~al.}(1995)\citenamefont {Stock},
  \citenamefont {Woywod}, \citenamefont {Domcke}, \citenamefont {Swinney},\
  and\ \citenamefont {Hudson}}]{Stock_Woywod:1995}%
  \BibitemOpen
  \bibfield  {author} {\bibinfo {author} {\bibfnamefont {G.}~\bibnamefont
  {Stock}}, \bibinfo {author} {\bibfnamefont {C.}~\bibnamefont {Woywod}},
  \bibinfo {author} {\bibfnamefont {W.}~\bibnamefont {Domcke}}, \bibinfo
  {author} {\bibfnamefont {T.}~\bibnamefont {Swinney}}, \ and\ \bibinfo
  {author} {\bibfnamefont {B.~S.}\ \bibnamefont {Hudson}},\ }\href {\doibase
  10.1063/1.470689} {\bibfield  {journal} {\bibinfo  {journal} {J.~Chem.\
  Phys.}\ }\textbf {\bibinfo {volume} {103}},\ \bibinfo {pages} {6851}
  (\bibinfo {year} {1995})}\BibitemShut {NoStop}%
\bibitem [{\citenamefont {Begu\v{s}i\'{c}}\ \emph {et~al.}(2018)\citenamefont
  {Begu\v{s}i\'{c}}, \citenamefont {Patoz}, \citenamefont {\v{S}ulc},\ and\
  \citenamefont {Van\'i\v{c}ek}}]{Begusic_Vanicek:2018}%
  \BibitemOpen
  \bibfield  {author} {\bibinfo {author} {\bibfnamefont {T.}~\bibnamefont
  {Begu\v{s}i\'{c}}}, \bibinfo {author} {\bibfnamefont {A.}~\bibnamefont
  {Patoz}}, \bibinfo {author} {\bibfnamefont {M.}~\bibnamefont {\v{S}ulc}}, \
  and\ \bibinfo {author} {\bibfnamefont {J.}~\bibnamefont {Van\'i\v{c}ek}},\
  }\href {\doibase https://doi.org/10.1016/j.chemphys.2018.08.003} {\bibfield
  {journal} {\bibinfo  {journal} {Chem.\ Phys.}\ }\textbf {\bibinfo {volume}
  {515}},\ \bibinfo {pages} {152} (\bibinfo {year} {2018})}\BibitemShut
  {NoStop}%
\bibitem [{\citenamefont {Ben-Nun}, \citenamefont {Quenneville},\ and\
  \citenamefont {Mart\'{\i}nez}(2000)}]{Ben-Nun_Martinez:2000}%
  \BibitemOpen
  \bibfield  {author} {\bibinfo {author} {\bibfnamefont {M.}~\bibnamefont
  {Ben-Nun}}, \bibinfo {author} {\bibfnamefont {J.}~\bibnamefont
  {Quenneville}}, \ and\ \bibinfo {author} {\bibfnamefont {T.~J.}\ \bibnamefont
  {Mart\'{\i}nez}},\ }\href {\doibase 10.1021/jp994174i} {\bibfield  {journal}
  {\bibinfo  {journal} {J.~Phys.\ Chem.~A}\ }\textbf {\bibinfo {volume}
  {104}},\ \bibinfo {pages} {5161} (\bibinfo {year} {2000})}\BibitemShut
  {NoStop}%
\bibitem [{\citenamefont {Bircher}\ \emph {et~al.}(2017)\citenamefont
  {Bircher}, \citenamefont {Liberatore}, \citenamefont {Browning},
  \citenamefont {Brickel}, \citenamefont {Hofmann}, \citenamefont {Patoz},
  \citenamefont {Unke}, \citenamefont {Zimmermann}, \citenamefont {Chergui},
  \citenamefont {Hamm}, \citenamefont {Keller}, \citenamefont {Meuwly},
  \citenamefont {Woerner}, \citenamefont {Van{\'{i}}{\v{c}}ek},\ and\
  \citenamefont {Rothlisberger}}]{Bircher_Rothlisberger:2017}%
  \BibitemOpen
  \bibfield  {author} {\bibinfo {author} {\bibfnamefont {M.~P.}\ \bibnamefont
  {Bircher}}, \bibinfo {author} {\bibfnamefont {E.}~\bibnamefont {Liberatore}},
  \bibinfo {author} {\bibfnamefont {N.~J.}\ \bibnamefont {Browning}}, \bibinfo
  {author} {\bibfnamefont {S.}~\bibnamefont {Brickel}}, \bibinfo {author}
  {\bibfnamefont {C.}~\bibnamefont {Hofmann}}, \bibinfo {author} {\bibfnamefont
  {A.}~\bibnamefont {Patoz}}, \bibinfo {author} {\bibfnamefont {O.~T.}\
  \bibnamefont {Unke}}, \bibinfo {author} {\bibfnamefont {T.}~\bibnamefont
  {Zimmermann}}, \bibinfo {author} {\bibfnamefont {M.}~\bibnamefont {Chergui}},
  \bibinfo {author} {\bibfnamefont {P.}~\bibnamefont {Hamm}}, \bibinfo {author}
  {\bibfnamefont {U.}~\bibnamefont {Keller}}, \bibinfo {author} {\bibfnamefont
  {M.}~\bibnamefont {Meuwly}}, \bibinfo {author} {\bibfnamefont {H.~J.}\
  \bibnamefont {Woerner}}, \bibinfo {author} {\bibfnamefont {J.}~\bibnamefont
  {Van{\'{i}}{\v{c}}ek}}, \ and\ \bibinfo {author} {\bibfnamefont
  {U.}~\bibnamefont {Rothlisberger}},\ }\href {\doibase 10.1063/1.4996816}
  {\bibfield  {journal} {\bibinfo  {journal} {Struct. Dyn.}\ }\textbf {\bibinfo
  {volume} {4}},\ \bibinfo {pages} {061510} (\bibinfo {year}
  {2017})}\BibitemShut {NoStop}%
\bibitem [{\citenamefont {Feit}, \citenamefont {Fleck},\ and\ \citenamefont
  {Steiger}(1982)}]{Feit_Steiger:1982}%
  \BibitemOpen
  \bibfield  {author} {\bibinfo {author} {\bibfnamefont {M.~D.}\ \bibnamefont
  {Feit}}, \bibinfo {author} {\bibfnamefont {J.~A.}\ \bibnamefont {Fleck},
  \bibfnamefont {{Jr.}}}, \ and\ \bibinfo {author} {\bibfnamefont
  {A.}~\bibnamefont {Steiger}},\ }\href@noop {} {\bibfield  {journal} {\bibinfo
   {journal} {J.~Comp.\ Phys.}\ }\textbf {\bibinfo {volume} {47}},\ \bibinfo
  {pages} {412} (\bibinfo {year} {1982})}\BibitemShut {NoStop}%
\bibitem [{\citenamefont {Tannor}(2007)}]{book_Tannor:2007}%
  \BibitemOpen
  \bibfield  {author} {\bibinfo {author} {\bibfnamefont {D.~J.}\ \bibnamefont
  {Tannor}},\ }\href@noop {} {\emph {\bibinfo {title} {Introduction to Quantum
  Mechanics: A Time-Dependent Perspective}}}\ (\bibinfo  {publisher}
  {University Science Books},\ \bibinfo {address} {Sausalito},\ \bibinfo {year}
  {2007})\BibitemShut {NoStop}%
\bibitem [{\citenamefont {Lubich}(2008)}]{book_Lubich:2008}%
  \BibitemOpen
  \bibfield  {author} {\bibinfo {author} {\bibfnamefont {C.}~\bibnamefont
  {Lubich}},\ }\href@noop {} {\emph {\bibinfo {title} {From Quantum to
  Classical Molecular Dynamics: Reduced Models and Numerical Analysis}}},\
  \bibinfo {edition} {12th}\ ed.\ (\bibinfo  {publisher} {European Mathematical
  Society},\ \bibinfo {year} {2008})\BibitemShut {NoStop}%
\bibitem [{\citenamefont {Hairer}, \citenamefont {Lubich},\ and\ \citenamefont
  {Wanner}(2006)}]{book_Hairer_Wanner:2006}%
  \BibitemOpen
  \bibfield  {author} {\bibinfo {author} {\bibfnamefont {E.}~\bibnamefont
  {Hairer}}, \bibinfo {author} {\bibfnamefont {C.}~\bibnamefont {Lubich}}, \
  and\ \bibinfo {author} {\bibfnamefont {G.}~\bibnamefont {Wanner}},\ }\href
  {http://books.google.ch/books/about/Geometric_Numerical_Integration.html?id=T1TaNRLmZv8C&redir_esc=y}
  {\emph {\bibinfo {title} {Geometric Numerical Integration:
  Structure-Preserving Algorithms for Ordinary Differential Equations}}}\
  (\bibinfo  {publisher} {Springer Berlin Heidelberg New York},\ \bibinfo
  {year} {2006})\BibitemShut {NoStop}%
\bibitem [{\citenamefont {Roulet}, \citenamefont {Choi},\ and\ \citenamefont
  {Van\'{i}\v{c}ek}(2019)}]{Roulet_Vanicek:2019}%
  \BibitemOpen
  \bibfield  {author} {\bibinfo {author} {\bibfnamefont {J.}~\bibnamefont
  {Roulet}}, \bibinfo {author} {\bibfnamefont {S.}~\bibnamefont {Choi}}, \ and\
  \bibinfo {author} {\bibfnamefont {J.}~\bibnamefont {Van\'{i}\v{c}ek}},\
  }\href {\doibase 10.1063/1.5094046} {\bibfield  {journal} {\bibinfo
  {journal} {J.~Chem.\ Phys.}\ }\textbf {\bibinfo {volume} {150}},\ \bibinfo
  {pages} {204113} (\bibinfo {year} {2019})}\BibitemShut {NoStop}%
\bibitem [{\citenamefont {Kosloff}\ and\ \citenamefont
  {Kosloff}(1983)}]{Kosloff_Kosloff:1983}%
  \BibitemOpen
  \bibfield  {author} {\bibinfo {author} {\bibfnamefont {D.}~\bibnamefont
  {Kosloff}}\ and\ \bibinfo {author} {\bibfnamefont {R.}~\bibnamefont
  {Kosloff}},\ }\href {\doibase 10.1016/0021-9991(83)90015-3} {\bibfield
  {journal} {\bibinfo  {journal} {J.~Comp.\ Phys.}\ }\textbf {\bibinfo {volume}
  {52}},\ \bibinfo {pages} {35} (\bibinfo {year} {1983})}\BibitemShut {NoStop}%
\bibitem [{\citenamefont {Meyer}, \citenamefont {Manthe},\ and\ \citenamefont
  {Cederbaum}(1990)}]{Meyer_Cederbaum:1990}%
  \BibitemOpen
  \bibfield  {author} {\bibinfo {author} {\bibfnamefont {H.-D.}\ \bibnamefont
  {Meyer}}, \bibinfo {author} {\bibfnamefont {U.}~\bibnamefont {Manthe}}, \
  and\ \bibinfo {author} {\bibfnamefont {L.~S.}\ \bibnamefont {Cederbaum}},\
  }\href {\doibase 10.1016/0009-2614(90)87014-I} {\bibfield  {journal}
  {\bibinfo  {journal} {Chem.\ Phys.\ Lett.}\ }\textbf {\bibinfo {volume}
  {165}},\ \bibinfo {pages} {73} (\bibinfo {year} {1990})}\BibitemShut
  {NoStop}%
\bibitem [{\citenamefont {Manthe}, \citenamefont {Meyer},\ and\ \citenamefont
  {Cederbaum}(1992)}]{Manthe_Cederbaum:1992}%
  \BibitemOpen
  \bibfield  {author} {\bibinfo {author} {\bibfnamefont {U.}~\bibnamefont
  {Manthe}}, \bibinfo {author} {\bibfnamefont {H.}~\bibnamefont {Meyer}}, \
  and\ \bibinfo {author} {\bibfnamefont {L.~S.}\ \bibnamefont {Cederbaum}},\
  }\href {\doibase 10.1063/1.463007} {\bibfield  {journal} {\bibinfo  {journal}
  {J.~Chem.\ Phys.}\ }\textbf {\bibinfo {volume} {97}},\ \bibinfo {pages}
  {3199} (\bibinfo {year} {1992})}\BibitemShut {NoStop}%
\bibitem [{\citenamefont {Worth}\ \emph {et~al.}(2008)\citenamefont {Worth},
  \citenamefont {Meyer}, \citenamefont {K\"{o}ppel}, \citenamefont
  {Cederbaum},\ and\ \citenamefont {Burghardt}}]{Worth_Burghardt:2008}%
  \BibitemOpen
  \bibfield  {author} {\bibinfo {author} {\bibfnamefont {G.~A.}\ \bibnamefont
  {Worth}}, \bibinfo {author} {\bibfnamefont {H.-D.}\ \bibnamefont {Meyer}},
  \bibinfo {author} {\bibfnamefont {H.}~\bibnamefont {K\"{o}ppel}}, \bibinfo
  {author} {\bibfnamefont {L.~S.}\ \bibnamefont {Cederbaum}}, \ and\ \bibinfo
  {author} {\bibfnamefont {I.}~\bibnamefont {Burghardt}},\ }\href {\doibase
  10.1080/01442350802137656} {\bibfield  {journal} {\bibinfo  {journal} {Int.\
  Rev.\ Phys.\ Chem.}\ }\textbf {\bibinfo {volume} {27}},\ \bibinfo {pages}
  {569} (\bibinfo {year} {2008})}\BibitemShut {NoStop}%
\bibitem [{\citenamefont {Wang}\ and\ \citenamefont
  {Thoss}(2003)}]{Wang_Thoss:2003}%
  \BibitemOpen
  \bibfield  {author} {\bibinfo {author} {\bibfnamefont {H.}~\bibnamefont
  {Wang}}\ and\ \bibinfo {author} {\bibfnamefont {M.}~\bibnamefont {Thoss}},\
  }\href {\doibase 10.1063/1.1580111} {\bibfield  {journal} {\bibinfo
  {journal} {J.~Chem.\ Phys.}\ }\textbf {\bibinfo {volume} {119}},\ \bibinfo
  {pages} {1289} (\bibinfo {year} {2003})}\BibitemShut {NoStop}%
\bibitem [{\citenamefont {Wodraszka}\ and\ \citenamefont
  {Carrington}(2019)}]{Wodraszka_Carrington:2019}%
  \BibitemOpen
  \bibfield  {author} {\bibinfo {author} {\bibfnamefont {R.}~\bibnamefont
  {Wodraszka}}\ and\ \bibinfo {author} {\bibfnamefont {T.}~\bibnamefont
  {Carrington}},\ }\href {\doibase 10.1063/1.5093317} {\bibfield  {journal}
  {\bibinfo  {journal} {J.~Chem.\ Phys.}\ }\textbf {\bibinfo {volume} {150}},\
  \bibinfo {pages} {154108} (\bibinfo {year} {2019})}\BibitemShut {NoStop}%
\bibitem [{\citenamefont {Wodraszka}\ and\ \citenamefont
  {Carrington}(2016)}]{Wodraszka_Carrington:2016}%
  \BibitemOpen
  \bibfield  {author} {\bibinfo {author} {\bibfnamefont {R.}~\bibnamefont
  {Wodraszka}}\ and\ \bibinfo {author} {\bibfnamefont {T.}~\bibnamefont
  {Carrington}},\ }\href {\doibase 10.1063/1.4959228} {\bibfield  {journal}
  {\bibinfo  {journal} {J.~Chem.\ Phys.}\ }\textbf {\bibinfo {volume} {145}},\
  \bibinfo {pages} {044110} (\bibinfo {year} {2016})}\BibitemShut {NoStop}%
\bibitem [{\citenamefont {Wodraszka}\ and\ \citenamefont
  {Carrington}(2018)}]{Wodraszka_Carrington:2018}%
  \BibitemOpen
  \bibfield  {author} {\bibinfo {author} {\bibfnamefont {R.}~\bibnamefont
  {Wodraszka}}\ and\ \bibinfo {author} {\bibfnamefont {T.}~\bibnamefont
  {Carrington}},\ }\href {\doibase 10.1063/1.5018793} {\bibfield  {journal}
  {\bibinfo  {journal} {J.~Chem.\ Phys.}\ }\textbf {\bibinfo {volume} {148}},\
  \bibinfo {pages} {044115} (\bibinfo {year} {2018})}\BibitemShut {NoStop}%
\bibitem [{\citenamefont {Davis}\ and\ \citenamefont
  {Heller}(1979)}]{Davis_Heller:1979}%
  \BibitemOpen
  \bibfield  {author} {\bibinfo {author} {\bibfnamefont {M.~J.}\ \bibnamefont
  {Davis}}\ and\ \bibinfo {author} {\bibfnamefont {E.~J.}\ \bibnamefont
  {Heller}},\ }\href {\doibase 10.1063/1.438727} {\bibfield  {journal}
  {\bibinfo  {journal} {J.~Chem.\ Phys.}\ }\textbf {\bibinfo {volume} {71}},\
  \bibinfo {pages} {3383} (\bibinfo {year} {1979})}\BibitemShut {NoStop}%
\bibitem [{\citenamefont {Shimshovitz}\ and\ \citenamefont
  {Tannor}(2012)}]{Shimshovitz_Tannor:2012}%
  \BibitemOpen
  \bibfield  {author} {\bibinfo {author} {\bibfnamefont {A.}~\bibnamefont
  {Shimshovitz}}\ and\ \bibinfo {author} {\bibfnamefont {D.~J.}\ \bibnamefont
  {Tannor}},\ }\href {\doibase 10.1103/PhysRevLett.109.070402} {\bibfield
  {journal} {\bibinfo  {journal} {Phys.\ Rev.\ Lett.}\ }\textbf {\bibinfo
  {volume} {109}},\ \bibinfo {pages} {070402} (\bibinfo {year}
  {2012})}\BibitemShut {NoStop}%
\bibitem [{\citenamefont {Halverson}\ and\ \citenamefont
  {Poirier}(2012)}]{Halverson_Poirier:2012}%
  \BibitemOpen
  \bibfield  {author} {\bibinfo {author} {\bibfnamefont {T.}~\bibnamefont
  {Halverson}}\ and\ \bibinfo {author} {\bibfnamefont {B.}~\bibnamefont
  {Poirier}},\ }\href {\doibase 10.1063/1.4769402} {\bibfield  {journal}
  {\bibinfo  {journal} {J.~Chem.\ Phys.}\ }\textbf {\bibinfo {volume} {137}},\
  \bibinfo {pages} {224101} (\bibinfo {year} {2012})}\BibitemShut {NoStop}%
\bibitem [{\citenamefont {Arai}\ \emph {et~al.}(2018)\citenamefont {Arai},
  \citenamefont {Suzuki}, \citenamefont {Kanno},\ and\ \citenamefont
  {Kono}}]{Arai_Kono:2018}%
  \BibitemOpen
  \bibfield  {author} {\bibinfo {author} {\bibfnamefont {Y.}~\bibnamefont
  {Arai}}, \bibinfo {author} {\bibfnamefont {K.}~\bibnamefont {Suzuki}},
  \bibinfo {author} {\bibfnamefont {M.}~\bibnamefont {Kanno}}, \ and\ \bibinfo
  {author} {\bibfnamefont {H.}~\bibnamefont {Kono}},\ }\href {\doibase
  10.1016/j.cplett.2018.07.022} {\bibfield  {journal} {\bibinfo  {journal}
  {Chem.\ Phys.\ Lett.}\ }\textbf {\bibinfo {volume} {708}},\ \bibinfo {pages}
  {170 } (\bibinfo {year} {2018})}\BibitemShut {NoStop}%
\bibitem [{\citenamefont {Pettey}\ and\ \citenamefont
  {Wyatt}(2006)}]{Pettey_Wyatt:2006}%
  \BibitemOpen
  \bibfield  {author} {\bibinfo {author} {\bibfnamefont {L.~R.}\ \bibnamefont
  {Pettey}}\ and\ \bibinfo {author} {\bibfnamefont {R.~E.}\ \bibnamefont
  {Wyatt}},\ }\href {\doibase https://doi.org/10.1016/j.cplett.2006.04.081}
  {\bibfield  {journal} {\bibinfo  {journal} {Chem.\ Phys.\ Lett.}\ }\textbf
  {\bibinfo {volume} {424}},\ \bibinfo {pages} {443 } (\bibinfo {year}
  {2006})}\BibitemShut {NoStop}%
\bibitem [{\citenamefont {Lee}\ and\ \citenamefont
  {Chou}(2018)}]{Lee_Chou:2018}%
  \BibitemOpen
  \bibfield  {author} {\bibinfo {author} {\bibfnamefont {T.-Y.}\ \bibnamefont
  {Lee}}\ and\ \bibinfo {author} {\bibfnamefont {C.-C.}\ \bibnamefont {Chou}},\
  }\href {\doibase 10.1021/acs.jpca.7b11932} {\bibfield  {journal} {\bibinfo
  {journal} {J.~Phys.\ Chem.~A}\ }\textbf {\bibinfo {volume} {122}},\ \bibinfo
  {pages} {1451} (\bibinfo {year} {2018})}\BibitemShut {NoStop}%
\bibitem [{\citenamefont {Avila}\ and\ \citenamefont
  {Carrington~Jr}(2009)}]{Avila_Carrington:2009}%
  \BibitemOpen
  \bibfield  {author} {\bibinfo {author} {\bibfnamefont {G.}~\bibnamefont
  {Avila}}\ and\ \bibinfo {author} {\bibfnamefont {T.}~\bibnamefont
  {Carrington~Jr}},\ }\href {\doibase 10.1063/1.3246593} {\bibfield  {journal}
  {\bibinfo  {journal} {J.~Chem.\ Phys.}\ }\textbf {\bibinfo {volume} {131}},\
  \bibinfo {pages} {174103} (\bibinfo {year} {2009})}\BibitemShut {NoStop}%
\bibitem [{\citenamefont {Gradinaru}(2008)}]{Gradinaru:2008}%
  \BibitemOpen
  \bibfield  {author} {\bibinfo {author} {\bibfnamefont {V.}~\bibnamefont
  {Gradinaru}},\ }\href {\doibase 10.1137/050629823} {\bibfield  {journal}
  {\bibinfo  {journal} {SIAM J.~Num.\ Analysis}\ }\textbf {\bibinfo {volume}
  {46}},\ \bibinfo {pages} {103} (\bibinfo {year} {2008})}\BibitemShut
  {NoStop}%
\bibitem [{\citenamefont {Lauvergnat}\ and\ \citenamefont
  {Nauts}(2014)}]{Lauvergnat_Nauts:2014}%
  \BibitemOpen
  \bibfield  {author} {\bibinfo {author} {\bibfnamefont {D.}~\bibnamefont
  {Lauvergnat}}\ and\ \bibinfo {author} {\bibfnamefont {A.}~\bibnamefont
  {Nauts}},\ }\href {\doibase 10.1016/j.saa.2013.05.068} {\bibfield  {journal}
  {\bibinfo  {journal} {Spectrochim. Acta A}\ }\textbf {\bibinfo {volume}
  {119}},\ \bibinfo {pages} {18 } (\bibinfo {year} {2014})}\BibitemShut
  {NoStop}%
\bibitem [{\citenamefont {Smolyak}(1963)}]{Smolyak:1963}%
  \BibitemOpen
  \bibfield  {author} {\bibinfo {author} {\bibfnamefont {S.~A.}\ \bibnamefont
  {Smolyak}},\ }\href@noop {} {\bibfield  {journal} {\bibinfo  {journal} {Dokl.
  Akad. Nauk SSSR}\ }\textbf {\bibinfo {volume} {148}},\ \bibinfo {pages}
  {1042} (\bibinfo {year} {1963})}\BibitemShut {NoStop}%
\bibitem [{\citenamefont {Thompson}, \citenamefont {Warsi},\ and\ \citenamefont
  {Mastin}(1997)}]{book_Thomson_Mastin:1997}%
  \BibitemOpen
  \bibfield  {author} {\bibinfo {author} {\bibfnamefont {J.~F.}\ \bibnamefont
  {Thompson}}, \bibinfo {author} {\bibfnamefont {Z.~U.}\ \bibnamefont {Warsi}},
  \ and\ \bibinfo {author} {\bibfnamefont {C.~W.}\ \bibnamefont {Mastin}},\
  }\href@noop {} {\emph {\bibinfo {title} {Numerical grid generation:
  foundations and applications}}},\ Vol.~\bibinfo {volume} {45}\ (\bibinfo
  {publisher} {North-holland Amsterdam},\ \bibinfo {year} {1997})\
  Chap.~\bibinfo {chapter} {11}\BibitemShut {NoStop}%
\bibitem [{\citenamefont {Lopreore}\ and\ \citenamefont
  {Wyatt}(1999)}]{Lopreore_Wyatt:1999}%
  \BibitemOpen
  \bibfield  {author} {\bibinfo {author} {\bibfnamefont {C.~L.}\ \bibnamefont
  {Lopreore}}\ and\ \bibinfo {author} {\bibfnamefont {R.~E.}\ \bibnamefont
  {Wyatt}},\ }\href {\doibase 10.1103/PhysRevLett.82.5190} {\bibfield
  {journal} {\bibinfo  {journal} {Phys.\ Rev.\ Lett.}\ }\textbf {\bibinfo
  {volume} {82}},\ \bibinfo {pages} {5190} (\bibinfo {year}
  {1999})}\BibitemShut {NoStop}%
\bibitem [{\citenamefont {Wyatt}(1999)}]{Wyatt:1999}%
  \BibitemOpen
  \bibfield  {author} {\bibinfo {author} {\bibfnamefont {R.~E.}\ \bibnamefont
  {Wyatt}},\ }\href {\doibase 10.1063/1.479205} {\bibfield  {journal} {\bibinfo
   {journal} {J.~Chem.\ Phys.}\ }\textbf {\bibinfo {volume} {111}},\ \bibinfo
  {pages} {4406} (\bibinfo {year} {1999})}\BibitemShut {NoStop}%
\bibitem [{\citenamefont {Wyatt}\ and\ \citenamefont
  {Bittner}(2000)}]{Wyatt_Bittner:2000}%
  \BibitemOpen
  \bibfield  {author} {\bibinfo {author} {\bibfnamefont {R.~E.}\ \bibnamefont
  {Wyatt}}\ and\ \bibinfo {author} {\bibfnamefont {E.~R.}\ \bibnamefont
  {Bittner}},\ }\href {\doibase 10.1063/1.1319988} {\bibfield  {journal}
  {\bibinfo  {journal} {J.~Chem.\ Phys.}\ }\textbf {\bibinfo {volume} {113}},\
  \bibinfo {pages} {8898} (\bibinfo {year} {2000})}\BibitemShut {NoStop}%
\bibitem [{\citenamefont {Wyatt}(2002)}]{Wyatt:2002}%
  \BibitemOpen
  \bibfield  {author} {\bibinfo {author} {\bibfnamefont {R.~E.}\ \bibnamefont
  {Wyatt}},\ }\href {\doibase 10.1063/1.1517045} {\bibfield  {journal}
  {\bibinfo  {journal} {J.~Chem.\ Phys.}\ }\textbf {\bibinfo {volume} {117}},\
  \bibinfo {pages} {9569} (\bibinfo {year} {2002})}\BibitemShut {NoStop}%
\bibitem [{\citenamefont {Hughes}\ and\ \citenamefont
  {Wyatt}(2002)}]{Hughes_Wyatt:2002}%
  \BibitemOpen
  \bibfield  {author} {\bibinfo {author} {\bibfnamefont {K.~H.}\ \bibnamefont
  {Hughes}}\ and\ \bibinfo {author} {\bibfnamefont {R.~E.}\ \bibnamefont
  {Wyatt}},\ }\href {\doibase 10.1016/S0009-2614(02)01654-8} {\bibfield
  {journal} {\bibinfo  {journal} {Chem.\ Phys.\ Lett.}\ }\textbf {\bibinfo
  {volume} {366}},\ \bibinfo {pages} {336} (\bibinfo {year}
  {2002})}\BibitemShut {NoStop}%
\bibitem [{\citenamefont {Lu}\ and\ \citenamefont
  {Bandrauk}(2001)}]{Lu_Bandrauk:2001}%
  \BibitemOpen
  \bibfield  {author} {\bibinfo {author} {\bibfnamefont {H.}~\bibnamefont
  {Lu}}\ and\ \bibinfo {author} {\bibfnamefont {A.~D.}\ \bibnamefont
  {Bandrauk}},\ }\href {\doibase 10.1063/1.1383033} {\bibfield  {journal}
  {\bibinfo  {journal} {J.~Chem.\ Phys.}\ }\textbf {\bibinfo {volume} {115}},\
  \bibinfo {pages} {1670} (\bibinfo {year} {2001})}\BibitemShut {NoStop}%
\bibitem [{\citenamefont {Sim}\ and\ \citenamefont
  {Makri}(1995)}]{Sim_Makri:1995}%
  \BibitemOpen
  \bibfield  {author} {\bibinfo {author} {\bibfnamefont {E.}~\bibnamefont
  {Sim}}\ and\ \bibinfo {author} {\bibfnamefont {N.}~\bibnamefont {Makri}},\
  }\href {\doibase 10.1063/1.469293} {\bibfield  {journal} {\bibinfo  {journal}
  {J.~Chem.\ Phys.}\ }\textbf {\bibinfo {volume} {102}},\ \bibinfo {pages}
  {5616} (\bibinfo {year} {1995})}\BibitemShut {NoStop}%
\bibitem [{\citenamefont {Adhikari}\ and\ \citenamefont
  {Billing}(2000)}]{Adhikari_Billing:2000}%
  \BibitemOpen
  \bibfield  {author} {\bibinfo {author} {\bibfnamefont {S.}~\bibnamefont
  {Adhikari}}\ and\ \bibinfo {author} {\bibfnamefont {G.~D.}\ \bibnamefont
  {Billing}},\ }\href {\doibase 10.1063/1.481959} {\bibfield  {journal}
  {\bibinfo  {journal} {J.~Chem.\ Phys.}\ }\textbf {\bibinfo {volume} {113}},\
  \bibinfo {pages} {1409} (\bibinfo {year} {2000})}\BibitemShut {NoStop}%
\bibitem [{\citenamefont {Lill}, \citenamefont {Parker},\ and\ \citenamefont
  {Light}(1982)}]{Lill_Light:1982}%
  \BibitemOpen
  \bibfield  {author} {\bibinfo {author} {\bibfnamefont {J.}~\bibnamefont
  {Lill}}, \bibinfo {author} {\bibfnamefont {G.}~\bibnamefont {Parker}}, \ and\
  \bibinfo {author} {\bibfnamefont {J.}~\bibnamefont {Light}},\ }\href
  {\doibase https://doi.org/10.1016/0009-2614(82)83051-0} {\bibfield  {journal}
  {\bibinfo  {journal} {Chem.\ Phys.\ Lett.}\ }\textbf {\bibinfo {volume}
  {89}},\ \bibinfo {pages} {483 } (\bibinfo {year} {1982})}\BibitemShut
  {NoStop}%
\bibitem [{\citenamefont {Light}\ and\ \citenamefont
  {Carrington~Jr.}(2007)}]{Light_Carrington:2007}%
  \BibitemOpen
  \bibfield  {author} {\bibinfo {author} {\bibfnamefont {J.~C.}\ \bibnamefont
  {Light}}\ and\ \bibinfo {author} {\bibfnamefont {T.}~\bibnamefont
  {Carrington~Jr.}},\ }\enquote {\bibinfo {title} {Discrete-variable
  representations and their utilization},}\ in\ \href {\doibase
  10.1002/9780470141731.ch4} {\emph {\bibinfo {booktitle} {Advances in Chemical
  Physics}}}\ (\bibinfo  {publisher} {John Wiley \& Sons, Ltd},\ \bibinfo
  {year} {2007})\ pp.\ \bibinfo {pages} {263--310}\BibitemShut {NoStop}%
\bibitem [{\citenamefont {Bramley}\ and\ \citenamefont
  {Carrington}(1993)}]{Bramley_Carrington:1993}%
  \BibitemOpen
  \bibfield  {author} {\bibinfo {author} {\bibfnamefont {M.~J.}\ \bibnamefont
  {Bramley}}\ and\ \bibinfo {author} {\bibfnamefont {T.}~\bibnamefont
  {Carrington}},\ }\href {\doibase 10.1063/1.465576} {\bibfield  {journal}
  {\bibinfo  {journal} {J.~Chem.\ Phys.}\ }\textbf {\bibinfo {volume} {99}},\
  \bibinfo {pages} {8519} (\bibinfo {year} {1993})}\BibitemShut {NoStop}%
\bibitem [{\citenamefont {Wang}\ and\ \citenamefont
  {Carrington}(2009)}]{Wang_Carrington:2009}%
  \BibitemOpen
  \bibfield  {author} {\bibinfo {author} {\bibfnamefont {X.-G.}\ \bibnamefont
  {Wang}}\ and\ \bibinfo {author} {\bibfnamefont {T.}~\bibnamefont
  {Carrington}},\ }\href {\doibase 10.1063/1.3077130} {\bibfield  {journal}
  {\bibinfo  {journal} {J.~Chem.\ Phys.}\ }\textbf {\bibinfo {volume} {130}},\
  \bibinfo {pages} {094101} (\bibinfo {year} {2009})}\BibitemShut {NoStop}%
\bibitem [{\citenamefont {Khan}\ \emph {et~al.}(2013)\citenamefont {Khan},
  \citenamefont {Sardar}, \citenamefont {Sahoo}, \citenamefont {Sarkar},\ and\
  \citenamefont {Adhikari}}]{Khan_Adhikari:2013}%
  \BibitemOpen
  \bibfield  {author} {\bibinfo {author} {\bibfnamefont {B.~A.}\ \bibnamefont
  {Khan}}, \bibinfo {author} {\bibfnamefont {S.}~\bibnamefont {Sardar}},
  \bibinfo {author} {\bibfnamefont {T.}~\bibnamefont {Sahoo}}, \bibinfo
  {author} {\bibfnamefont {P.}~\bibnamefont {Sarkar}}, \ and\ \bibinfo {author}
  {\bibfnamefont {S.}~\bibnamefont {Adhikari}},\ }\href {\doibase
  10.1142/S0219633613500429} {\bibfield  {journal} {\bibinfo  {journal} {J.
  Theor. Comput. Chem.}\ }\textbf {\bibinfo {volume} {12}},\ \bibinfo {pages}
  {1350042} (\bibinfo {year} {2013})}\BibitemShut {NoStop}%
\bibitem [{\citenamefont {Khan}\ \emph {et~al.}(2014)\citenamefont {Khan},
  \citenamefont {Sardar}, \citenamefont {Sarkar},\ and\ \citenamefont
  {Adhikari}}]{Khan_Adhikari:2014}%
  \BibitemOpen
  \bibfield  {author} {\bibinfo {author} {\bibfnamefont {B.~A.}\ \bibnamefont
  {Khan}}, \bibinfo {author} {\bibfnamefont {S.}~\bibnamefont {Sardar}},
  \bibinfo {author} {\bibfnamefont {P.}~\bibnamefont {Sarkar}}, \ and\ \bibinfo
  {author} {\bibfnamefont {S.}~\bibnamefont {Adhikari}},\ }\href {\doibase
  10.1021/jp507459m} {\bibfield  {journal} {\bibinfo  {journal} {J.~Phys.\
  Chem.~A}\ }\textbf {\bibinfo {volume} {118}},\ \bibinfo {pages} {11451}
  (\bibinfo {year} {2014})}\BibitemShut {NoStop}%
\bibitem [{\citenamefont {Puzari}, \citenamefont {Sarkar},\ and\ \citenamefont
  {Adhikari}(2005)}]{Puzari_Adhikari:2005}%
  \BibitemOpen
  \bibfield  {author} {\bibinfo {author} {\bibfnamefont {P.}~\bibnamefont
  {Puzari}}, \bibinfo {author} {\bibfnamefont {B.}~\bibnamefont {Sarkar}}, \
  and\ \bibinfo {author} {\bibfnamefont {S.}~\bibnamefont {Adhikari}},\ }\href
  {\doibase 10.1002/qua.20666} {\bibfield  {journal} {\bibinfo  {journal} {Int.
  J. Quantum Chem}\ }\textbf {\bibinfo {volume} {105}},\ \bibinfo {pages} {209}
  (\bibinfo {year} {2005})}\BibitemShut {NoStop}%
\bibitem [{\citenamefont {Mandal}\ \emph {et~al.}(2018)\citenamefont {Mandal},
  \citenamefont {Ghosh}, \citenamefont {Sardar},\ and\ \citenamefont
  {Adhikari}}]{Mandal_Adhikari:2018}%
  \BibitemOpen
  \bibfield  {author} {\bibinfo {author} {\bibfnamefont {S.}~\bibnamefont
  {Mandal}}, \bibinfo {author} {\bibfnamefont {S.}~\bibnamefont {Ghosh}},
  \bibinfo {author} {\bibfnamefont {S.}~\bibnamefont {Sardar}}, \ and\ \bibinfo
  {author} {\bibfnamefont {S.}~\bibnamefont {Adhikari}},\ }\href {\doibase
  10.1080/0144235X.2018.1548103} {\bibfield  {journal} {\bibinfo  {journal}
  {Int. Rev. Phys. Chem.}\ }\textbf {\bibinfo {volume} {37}},\ \bibinfo {pages}
  {607} (\bibinfo {year} {2018})}\BibitemShut {NoStop}%
\bibitem [{\citenamefont {Suzuki}(1990)}]{Suzuki:1990}%
  \BibitemOpen
  \bibfield  {author} {\bibinfo {author} {\bibfnamefont {M.}~\bibnamefont
  {Suzuki}},\ }\href {\doibase 10.1016/0375-9601(90)90962-n} {\bibfield
  {journal} {\bibinfo  {journal} {Phys.\ Lett.~A}\ }\textbf {\bibinfo {volume}
  {146}},\ \bibinfo {pages} {319} (\bibinfo {year} {1990})}\BibitemShut
  {NoStop}%
\bibitem [{\citenamefont {Yoshida}(1990)}]{Yoshida:1990}%
  \BibitemOpen
  \bibfield  {author} {\bibinfo {author} {\bibfnamefont {H.}~\bibnamefont
  {Yoshida}},\ }\href {\doibase 10.1016/0375-9601(90)90092-3} {\bibfield
  {journal} {\bibinfo  {journal} {Phys.\ Lett.~A}\ }\textbf {\bibinfo {volume}
  {150}},\ \bibinfo {pages} {262} (\bibinfo {year} {1990})}\BibitemShut
  {NoStop}%
\bibitem [{\citenamefont {Kahan}\ and\ \citenamefont
  {Li}(1997)}]{Kahan_Li:1997}%
  \BibitemOpen
  \bibfield  {author} {\bibinfo {author} {\bibfnamefont {W.}~\bibnamefont
  {Kahan}}\ and\ \bibinfo {author} {\bibfnamefont {R.-C.}\ \bibnamefont {Li}},\
  }\href {\doibase 10.1090/s0025-5718-97-00873-9} {\bibfield  {journal}
  {\bibinfo  {journal} {Math. Comput.}\ }\textbf {\bibinfo {volume} {66}},\
  \bibinfo {pages} {1089} (\bibinfo {year} {1997})}\BibitemShut {NoStop}%
\bibitem [{\citenamefont {Sofroniou}\ and\ \citenamefont
  {Spaletta}(2005)}]{Sofroniou_Spaletta:2005}%
  \BibitemOpen
  \bibfield  {author} {\bibinfo {author} {\bibfnamefont {M.}~\bibnamefont
  {Sofroniou}}\ and\ \bibinfo {author} {\bibfnamefont {G.}~\bibnamefont
  {Spaletta}},\ }\href {\doibase 10.1080/10556780500140664} {\bibfield
  {journal} {\bibinfo  {journal} {Optim. Method Softw.}\ }\textbf {\bibinfo
  {volume} {20}},\ \bibinfo {pages} {597} (\bibinfo {year} {2005})}\BibitemShut
  {NoStop}%
\bibitem [{\citenamefont {Ehrenfest}(1927)}]{Ehrenfest:1927}%
  \BibitemOpen
  \bibfield  {author} {\bibinfo {author} {\bibfnamefont {P.}~\bibnamefont
  {Ehrenfest}},\ }\href@noop {} {\bibfield  {journal} {\bibinfo  {journal} {Z.
  Phys}\ }\textbf {\bibinfo {volume} {45}},\ \bibinfo {pages} {455} (\bibinfo
  {year} {1927})}\BibitemShut {NoStop}%
\bibitem [{\citenamefont {Secrest}\ and\ \citenamefont
  {Johnson}(1966)}]{Secrest_Johnson:1966}%
  \BibitemOpen
  \bibfield  {author} {\bibinfo {author} {\bibfnamefont {D.}~\bibnamefont
  {Secrest}}\ and\ \bibinfo {author} {\bibfnamefont {B.~R.}\ \bibnamefont
  {Johnson}},\ }\href {\doibase 10.1063/1.1727537} {\bibfield  {journal}
  {\bibinfo  {journal} {J.~Chem.\ Phys.}\ }\textbf {\bibinfo {volume} {45}},\
  \bibinfo {pages} {4556} (\bibinfo {year} {1966})}\BibitemShut {NoStop}%
\bibitem [{\citenamefont {Campos-Mart{\'i}nez}\ and\ \citenamefont
  {Coalson}(1990)}]{Campos-Martinez_Coalson:1990}%
  \BibitemOpen
  \bibfield  {author} {\bibinfo {author} {\bibfnamefont {J.}~\bibnamefont
  {Campos-Mart{\'i}nez}}\ and\ \bibinfo {author} {\bibfnamefont {R.~D.}\
  \bibnamefont {Coalson}},\ }\href {\doibase 10.1063/1.458664} {\bibfield
  {journal} {\bibinfo  {journal} {J.~Chem.\ Phys.}\ }\textbf {\bibinfo {volume}
  {93}},\ \bibinfo {pages} {4740} (\bibinfo {year} {1990})}\BibitemShut
  {NoStop}%
\bibitem [{\citenamefont {Trotter}(1959)}]{Trotter:1959}%
  \BibitemOpen
  \bibfield  {author} {\bibinfo {author} {\bibfnamefont {H.~F.}\ \bibnamefont
  {Trotter}},\ }\href {\doibase 10.1090/S0002-9939-1959-0108732-6} {\bibfield
  {journal} {\bibinfo  {journal} {Proc.\ Amer.\ Math.\ Soc.}\ }\textbf
  {\bibinfo {volume} {10}},\ \bibinfo {pages} {545} (\bibinfo {year}
  {1959})}\BibitemShut {NoStop}%
\bibitem [{\citenamefont {Strang}(1968)}]{Strang:1968}%
  \BibitemOpen
  \bibfield  {author} {\bibinfo {author} {\bibfnamefont {G.}~\bibnamefont
  {Strang}},\ }\href {\doibase 10.1137/0705041} {\bibfield  {journal} {\bibinfo
   {journal} {SIAM J. Numer. Anal.}\ }\textbf {\bibinfo {volume} {5}},\
  \bibinfo {pages} {506} (\bibinfo {year} {1968})}\BibitemShut {NoStop}%
\bibitem [{\citenamefont {Leimkuhler}\ and\ \citenamefont
  {Reich}(2004)}]{book_Leimkuhler_Reich:2004}%
  \BibitemOpen
  \bibfield  {author} {\bibinfo {author} {\bibfnamefont {B.}~\bibnamefont
  {Leimkuhler}}\ and\ \bibinfo {author} {\bibfnamefont {S.}~\bibnamefont
  {Reich}},\ }\href
  {http://books.google.ch/books/about/Simulating_Hamiltonian_Dynamics.html?id=tpb-tnsZi5YC&redir_esc=y}
  {\emph {\bibinfo {title} {Simulating Hamiltonian Dynamics}}}\ (\bibinfo
  {publisher} {Cambridge University Press},\ \bibinfo {year}
  {2004})\BibitemShut {NoStop}%
\bibitem [{\citenamefont {Frigo}\ and\ \citenamefont
  {Johnson}(2005)}]{Frigo_Johnson:2005}%
  \BibitemOpen
  \bibfield  {author} {\bibinfo {author} {\bibfnamefont {M.}~\bibnamefont
  {Frigo}}\ and\ \bibinfo {author} {\bibfnamefont {S.~G.}\ \bibnamefont
  {Johnson}},\ }\href {\doibase 10.1109/JPROC.2004.840301} {\bibfield
  {journal} {\bibinfo  {journal} {Proc. IEEE}\ }\textbf {\bibinfo {volume}
  {93}},\ \bibinfo {pages} {216} (\bibinfo {year} {2005})}\BibitemShut
  {NoStop}%
\bibitem [{\citenamefont {Halmos}(1942)}]{book_Halmos:1942}%
  \BibitemOpen
  \bibfield  {author} {\bibinfo {author} {\bibfnamefont {P.~R.}\ \bibnamefont
  {Halmos}},\ }\href@noop {} {\emph {\bibinfo {title} {Finite dimensional
  vector spaces}}}\ (\bibinfo  {publisher} {Princeton University Press},\
  \bibinfo {year} {1942})\BibitemShut {NoStop}%
\bibitem [{\citenamefont {Lanczos}(1950)}]{Lanczos:1950}%
  \BibitemOpen
  \bibfield  {author} {\bibinfo {author} {\bibfnamefont {C.}~\bibnamefont
  {Lanczos}},\ }\href {\doibase 10.6028/jres.045.026} {\bibfield  {journal}
  {\bibinfo  {journal} {J. Res. Nat. Bur. Stand.}\ }\textbf {\bibinfo {volume}
  {45}},\ \bibinfo {pages} {255} (\bibinfo {year} {1950})}\BibitemShut
  {NoStop}%
\bibitem [{\citenamefont {Park}\ and\ \citenamefont
  {Light}(1986)}]{Park_Light:1986}%
  \BibitemOpen
  \bibfield  {author} {\bibinfo {author} {\bibfnamefont {T.~J.}\ \bibnamefont
  {Park}}\ and\ \bibinfo {author} {\bibfnamefont {J.~C.}\ \bibnamefont
  {Light}},\ }\href {\doibase 10.1063/1.451548} {\bibfield  {journal} {\bibinfo
   {journal} {J.~Chem.\ Phys.}\ }\textbf {\bibinfo {volume} {85}},\ \bibinfo
  {pages} {5870} (\bibinfo {year} {1986})}\BibitemShut {NoStop}%
\bibitem [{\citenamefont {Crank}\ and\ \citenamefont
  {Nicolson}(1947)}]{Crank_Nicolson:1947}%
  \BibitemOpen
  \bibfield  {author} {\bibinfo {author} {\bibfnamefont {J.}~\bibnamefont
  {Crank}}\ and\ \bibinfo {author} {\bibfnamefont {P.}~\bibnamefont
  {Nicolson}},\ }\href {\doibase 10.1017/s0305004100023197} {\bibfield
  {journal} {\bibinfo  {journal} {Math.\ Proc.\ Camb.\ Phil.\ Soc.}\ }\textbf
  {\bibinfo {volume} {43}},\ \bibinfo {pages} {50} (\bibinfo {year}
  {1947})}\BibitemShut {NoStop}%
\bibitem [{\citenamefont {McCullough}\ and\ \citenamefont
  {Wyatt}(1971)}]{McCullough_Wyatt:1971}%
  \BibitemOpen
  \bibfield  {author} {\bibinfo {author} {\bibfnamefont {E.~A.}\ \bibnamefont
  {McCullough}, \bibfnamefont {{Jr.}}}\ and\ \bibinfo {author} {\bibfnamefont
  {R.~E.}\ \bibnamefont {Wyatt}},\ }\href@noop {} {\bibfield  {journal}
  {\bibinfo  {journal} {J.~Chem.\ Phys.}\ }\textbf {\bibinfo {volume} {54}},\
  \bibinfo {pages} {3578} (\bibinfo {year} {1971})}\BibitemShut {NoStop}%
\bibitem [{\citenamefont {Choi}\ and\ \citenamefont
  {Van\'{i}\v{c}ek}(2019)}]{Choi_Vanicek:2019}%
  \BibitemOpen
  \bibfield  {author} {\bibinfo {author} {\bibfnamefont {S.}~\bibnamefont
  {Choi}}\ and\ \bibinfo {author} {\bibfnamefont {J.}~\bibnamefont
  {Van\'{i}\v{c}ek}},\ }\href {\doibase 10.1063/1.5092611} {\bibfield
  {journal} {\bibinfo  {journal} {J.~Chem.\ Phys.}\ }\textbf {\bibinfo {volume}
  {150}},\ \bibinfo {pages} {204112} (\bibinfo {year} {2019})}\BibitemShut
  {NoStop}%
\bibitem [{\citenamefont {Verlet}(1967)}]{Verlet:1967}%
  \BibitemOpen
  \bibfield  {author} {\bibinfo {author} {\bibfnamefont {L.}~\bibnamefont
  {Verlet}},\ }\href {\doibase 10.1103/PhysRev.159.98} {\bibfield  {journal}
  {\bibinfo  {journal} {Phys.\ Rev.}\ }\textbf {\bibinfo {volume} {159}},\
  \bibinfo {pages} {98} (\bibinfo {year} {1967})}\BibitemShut {NoStop}%
\bibitem [{\citenamefont {Frenkel}\ and\ \citenamefont
  {Smit}(2002)}]{Frenkel_Smit:2002}%
  \BibitemOpen
  \bibfield  {author} {\bibinfo {author} {\bibfnamefont {D.}~\bibnamefont
  {Frenkel}}\ and\ \bibinfo {author} {\bibfnamefont {B.}~\bibnamefont {Smit}},\
  }\href@noop {} {\emph {\bibinfo {title} {Understanding molecular
  simulation}}},\ \bibinfo {edition} {2nd}\ ed.\ (\bibinfo  {publisher}
  {Academic Press},\ \bibinfo {year} {2002})\BibitemShut {NoStop}%
\bibitem [{\citenamefont {Bhatia}\ and\ \citenamefont
  {Szeg{\"o}}(2006)}]{book_Bhatia_George:2006}%
  \BibitemOpen
  \bibfield  {author} {\bibinfo {author} {\bibfnamefont {N.~P.}\ \bibnamefont
  {Bhatia}}\ and\ \bibinfo {author} {\bibfnamefont {G.~P.}\ \bibnamefont
  {Szeg{\"o}}},\ }\href@noop {} {\emph {\bibinfo {title} {Dynamical systems:
  stability theory and applications}}},\ Vol.~\bibinfo {volume} {35}\ (\bibinfo
   {publisher} {Springer},\ \bibinfo {year} {2006})\BibitemShut {NoStop}%
\bibitem [{\citenamefont {Duschinsky}(1937)}]{Duschinsky:1937}%
  \BibitemOpen
  \bibfield  {author} {\bibinfo {author} {\bibfnamefont {F.}~\bibnamefont
  {Duschinsky}},\ }\href
  {http://www.ruhr-uni-bochum.de/pc2/mam/duschinsky{\_}translation.pdf}
  {\bibfield  {journal} {\bibinfo  {journal} {Acta Physicochim. U.R.S.S.}\
  }\textbf {\bibinfo {volume} {7}},\ \bibinfo {pages} {551} (\bibinfo {year}
  {1937})}\BibitemShut {NoStop}%
\bibitem [{\citenamefont {Nest}\ and\ \citenamefont
  {Meyer}(2002)}]{Nest_Meyer:2002}%
  \BibitemOpen
  \bibfield  {author} {\bibinfo {author} {\bibfnamefont {M.}~\bibnamefont
  {Nest}}\ and\ \bibinfo {author} {\bibfnamefont {H.-D.}\ \bibnamefont
  {Meyer}},\ }\href {\doibase 10.1063/1.1521129} {\bibfield  {journal}
  {\bibinfo  {journal} {The Journal of Chemical Physics}\ }\textbf {\bibinfo
  {volume} {117}},\ \bibinfo {pages} {10499} (\bibinfo {year}
  {2002})}\BibitemShut {NoStop}%
\bibitem [{\citenamefont {Tal‐Ezer}\ and\ \citenamefont
  {Kosloff}(1984)}]{Tal-Ezer_Kosloff:1984}%
  \BibitemOpen
  \bibfield  {author} {\bibinfo {author} {\bibfnamefont {H.}~\bibnamefont
  {Tal‐Ezer}}\ and\ \bibinfo {author} {\bibfnamefont {R.}~\bibnamefont
  {Kosloff}},\ }\href {\doibase 10.1063/1.448136} {\bibfield  {journal}
  {\bibinfo  {journal} {J.~Chem.\ Phys.}\ }\textbf {\bibinfo {volume} {81}},\
  \bibinfo {pages} {3967} (\bibinfo {year} {1984})}\BibitemShut {NoStop}%
\bibitem [{\citenamefont {Heller}(1975)}]{Heller:1975}%
  \BibitemOpen
  \bibfield  {author} {\bibinfo {author} {\bibfnamefont {E.~J.}\ \bibnamefont
  {Heller}},\ }\href {\doibase 10.1063/1.430620} {\bibfield  {journal}
  {\bibinfo  {journal} {J.~Chem.\ Phys.}\ }\textbf {\bibinfo {volume} {62}},\
  \bibinfo {pages} {1544} (\bibinfo {year} {1975})}\BibitemShut {NoStop}%
\bibitem [{\citenamefont {Wehrle}, \citenamefont {\v{S}ulc},\ and\
  \citenamefont {Van\'{i}\v{c}ek}(2014)}]{Wehrle_Vanicek:2014}%
  \BibitemOpen
  \bibfield  {author} {\bibinfo {author} {\bibfnamefont {M.}~\bibnamefont
  {Wehrle}}, \bibinfo {author} {\bibfnamefont {M.}~\bibnamefont {\v{S}ulc}}, \
  and\ \bibinfo {author} {\bibfnamefont {J.}~\bibnamefont {Van\'{i}\v{c}ek}},\
  }\href@noop {} {\bibfield  {journal} {\bibinfo  {journal} {J.~Chem.\ Phys.}\
  }\textbf {\bibinfo {volume} {140}},\ \bibinfo {pages} {244114} (\bibinfo
  {year} {2014})}\BibitemShut {NoStop}%
\bibitem [{\citenamefont {Begu\v{s}i\'{c}}, \citenamefont {Cordova},\ and\
  \citenamefont {Van{\'{i}}{\v{c}}ek}(2019)}]{Begusic_Vanicek:2019}%
  \BibitemOpen
  \bibfield  {author} {\bibinfo {author} {\bibfnamefont {T.}~\bibnamefont
  {Begu\v{s}i\'{c}}}, \bibinfo {author} {\bibfnamefont {M.}~\bibnamefont
  {Cordova}}, \ and\ \bibinfo {author} {\bibfnamefont {J.}~\bibnamefont
  {Van{\'{i}}{\v{c}}ek}},\ }\href {\doibase 10.1063/1.5090122} {\bibfield
  {journal} {\bibinfo  {journal} {J.~Chem.\ Phys.}\ }\textbf {\bibinfo {volume}
  {150}},\ \bibinfo {pages} {154117} (\bibinfo {year} {2019})}\BibitemShut
  {NoStop}%
\end{thebibliography}%

\end{document}